\RequirePackage{fix-cm}
\documentclass[twocolumn]{svjour3}       
\smartqed  
\usepackage{enumitem}
\usepackage{url}
\usepackage{algorithm}
\usepackage{algorithmicx}
\usepackage{algpseudocode}
\usepackage{latexsym}
\usepackage{color}
\usepackage{epsfig}
\usepackage{subfigure}
\usepackage{amsfonts}
\usepackage{amssymb}
\newcommand{\etal}[0]{{\it et al. }}

\usepackage{epstopdf}
\AppendGraphicsExtensions{-generated.png}
\usepackage{mathtools,etoolbox}
\begin{document}

\title{Automatic Delineation of Kidney Region in DCE-MRI
}

\titlerunning{Automatic Delineation of Kidney Region in DCE-MRI}        

\author{Santosh~Tirunagari, 
		Norman~Poh, 
		Kevin~Wells, 
		Miroslaw~Bober, 
		Isky~Gorden, 
        and David~Windridge.
}

\authorrunning{Tirunagari \etal} 

\institute{S. Tirunagari, N. Poh,  K. Wells and M. Bober are \at
              with Faculty of Engineering and Physical Sciences (FEPS)\\
	      University of Surrey\\
	      Guildford \\
              \email{S.tirunagari@surrey.ac.uk}           
           \and
           D.Windridge is with the Department of Computer Science, Middlesex University, London, UK. I. Gordon is with the UCL Institute of Child Health, 30 Guildford Street, London, WCIN lEH, UK.
}

\date{Received: date / Accepted: date}

\maketitle

\begin{abstract}
Delineation of the kidney region in dynamic contrast enhanced magnetic resonance Imaging (DCE-MRI) is required during post-acquisition analysis in order to quantify various aspects of renal function, such as filtration and perfusion or blood flow. However, this can be obfuscated by the Partial Volume Effect (PVE), caused due to mixing of any single voxel with two or more signal intensities from adjacent regions such as liver region and other tissues. To avoid this problem, firstly, a kidney region of interest (ROI) needs to be defined for the analysis. A clinician may choose to select a region avoiding edges where PV mixing is likely to be significant. However, this approach is time-consuming and labour intensive. To address this issue, we present Dynamic Mode Decomposition (DMD) coupled with thresholding and blob analysis as a framework for automatic delineation of the kidney region. This method is first validated on synthetically generated data with ground-truth available and then applied to ten healthy volunteers' kidney DCE-MRI datasets. We found that the result obtained from our proposed framework is comparable to that of a human expert. For example, while our result gives an average Root Mean Square Error (RMSE) of $0.0097$, the baseline achieves an average RMSE of $0.1196$ across the $10$ datasets. As a result, we conclude automatic modelling via DMD framework is a promising approach.

\end{abstract}

\section{Introduction}

\begin{figure*}
\centering
\begin{tabular}{cc}
\includegraphics[scale=0.3]{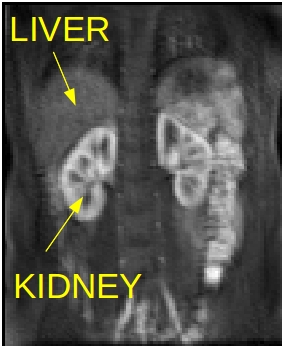} & \includegraphics[scale=0.7]{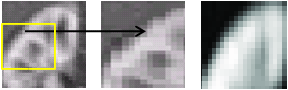}\\
(a) & (b)\\
\end{tabular}
\caption{(a) DCE-MR image of a healthy volunteer  where the kidney is contrast enhanced. Quantification of kidney function requires computation of mean intensities of the pixel values. However, the quantification is contaminated by the mixing of intensity values from liver region and background artefacts due to complex patient movements. (b) An example of blurring or partial volume effect using a segment of a DCE-MR image. The image on the middle shows a magnified region from the kidney-liver boundary. Clearly there is no definitive boundary between the tissues due to the limited resolution of the image, which indicates that the voxels may contain more than one signal intensity. The blurring effect, however, is not limited to the boundaries such as above since the kidney has a complex structure that consists of different tissues as shown in (a). The image on the right shows the delineation of kidney region by the proposed framework clearly suppressing the mixing from the liver region.}
\label{anatomy}
\end{figure*}

DCE-MRI renography involves injection of an MRI contrast agent into the blood stream for obtaining magnetic resonance images (MRI) of internal body structures such as kidney, liver and spleen as shown in Figure~\ref{anatomy}. The images are taken sequentially over the time, until the contrast agent is discharged. Delineation of the kidney region from these dynamic image sequence is required in order to quantify various aspects of renal function, such as filtration and perfusion or blood flow. However, Absolute quantification in DCE-MRI is often obfuscated by the issues including region of interest (ROI) definition, which may be influenced by the adjacent regions giving rise to the so called Partial Volume Effect (PVE). PVE is the contamination of any single voxel with two or more signal intensities from different tissues or regions (refer Figure~\ref{anatomy} for more explanation). The contamination is caused by the finite resolution of imaging systems that limits the higher frequencies i.e., fine details such as edges and structures. This results in producing inaccurate or blurred images especially in DCE-MRI where the spatial information is sacrificed for better temporal resolution since time is an important factor in the assessment of kidney function. The two primary issues concerning PVE are: (i) the action of the Point Spread function on the image data, and (2) the action of finite voxel sampling (comb function) and the integration of signal within the volume of the voxel (rect function). Further details, albeit for the same effects in 2D can be found in Yip et al, Phys Med Biol 2011, listed under my papers on the uni web page. 
Therefore, DCE-MRI is required to undergo several steps including: 1) suppression of PVE~\cite{gutierrez2010partial}, selection of kidney ROI~\cite{zollner2009assessment,rusinek2007performance} and a final stage of modelling kidney function by analysis of the resulting time-intensity plots~\cite{tofts2012precise}.

A template-based approach based on tissue classification proposed in~\cite{gutierrez2010partial} has demonstrated that PVE is a very significant issue in quantification of DCE-MRI studies. From the literature, eigen based methods such as Independent component analysis (ICA) and its variants have also been used for obtaining the kidney ROI~\cite{zollner2007assessment}. In addition, k-means clustering ~\cite{zollner2009assessment} has also been employed. Although these methods have shown improvements in the processing of the data, a complete automated solution has not been provided. The major issue with the aforementioned approaches is the need for manual or semi-automatic delineation of the target region, and the need for prior knowledge of the point spread function of the particular MRI sequence used, which will vary between particular sequences and vendors. In addition, this can be labour intensive, time-consuming and inefficient as the human expert has to examine the whole sequence of images to find the most suitable frame for delineation. Therefore, to address this problem in more holistic manner, we in this study, present a framework consisting of DMD, thresholding and blob analysis for automatic delineation of kidney ROI with the PVE compensation within the DCE-MRI sequence.

DMD was originally introduced in the area of computational fluid dynamics (CFD)~\cite{Schmid2}, specifically for analysing the sequential image data generated by non-linear complex fluid flows~\cite{Schmid1,Schmid2,Schmid3,tirunagarianalysis}. The DMD decomposes a given image sequence into several images, called dynamic modes. These modes essentially capture different large scale to small scale structures (sparse components) including a background structure (low rank model)~\cite{DBLP:journals/corr/GrosekK14,randDMD,tirunagari2016can}.  DMD has gained significant applications in various fields~\cite{6926317,2015arXiv150804487M,brunton2016extracting}, including for detecting spoof samples from facial authentication video datasets~\cite{tirunagari2015detection} and for detecting spoofed finger-vein images~\cite{tirunagari2015windowed}. The advantage of this method is its ability to identify regions of dominant motion in an image sequence in a completely data-driven manner without relying on any prior assumptions about the patterns of behaviour within the data. Therefore, it is thus potentially well-suited to analyse a wide variation of blood flow and filtration patterns seen in renography pathology.

\begin{figure}[h]
\centering
\begin{tabular}{c}
\includegraphics[scale=0.55]{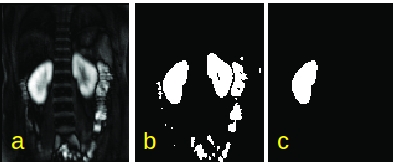}\\
\end{tabular}
\caption{(a) DMD mode 2 showing the pre-dominant kidney region by suppressing the background and liver regions. (b) Thresholded version of the DMD mode-2. (c) Selection of the kidney ROI using the blob analysis.}
\label{prelim-result}
\end{figure}

Our preliminary results show that DMD mode-2 highlights the kidney region due to the presence of injected contrast agent. Therefore, as shown in Figure~\ref{prelim-result}, this mode is then used for delineating the kidney region using thresholding and blob analysis.

Prior works directed towards tackling the issues related to Partial Volume Correction (PVE) and delineation of kidney ROI are often dealt separately using different methods. In particular, to our knowledge, there is no literature of using a single method that is able to tackle both of these issues simultaneously. Therefore, our first contribution is to introduce a single methodology that is based on DMD to tackle the aforementioned issues problem. Our second contribution is to Validate our technique for the first time on medical data with applications to DCE-MRI; and finally, our third contribution is in improving the understanding of the application through our DMD framework.

The remainder of this paper is organised as follows: in Section~\ref{method}, we consider the theory for DMD. Section~\ref{data} presents synthetically generated dataset as well as ten different DCE-MRI datasets used in this study. In Section~\ref{exp_res}  we present our experiments and results and finally, conclusions are drawn in Section~\ref{conc}.


\section{Methodology}
\label{method}

\begin{figure*}[!ht]
\centering
\includegraphics[scale=0.6]{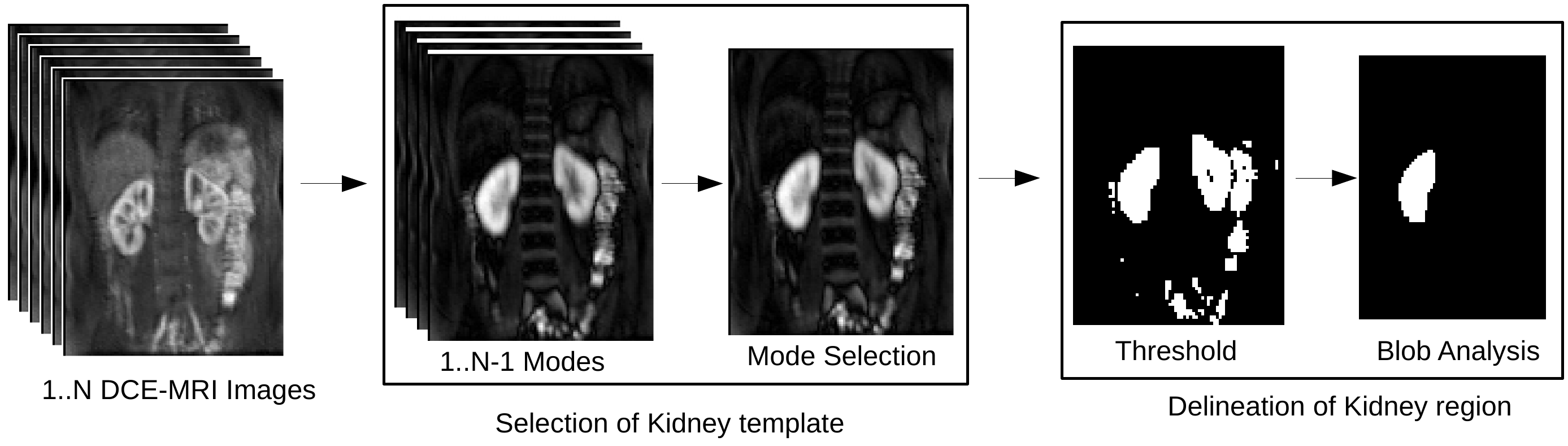}
\caption{Flow chart showing the steps involved in the methodological framework. 
First, a DCE-MRI sequence consisting of $N$ images is processed using the DMD algorithm in order to output $N-1$ dynamic mode images.
From which, we select a dynamic mode image corresponding to mode = $2$. Second, thresholding is performed on this mode and converted to binary image. After
binarization, the largest area with connected pixels is selected as template. Finally, the produced template is projected onto
DCE-MRI data to quantify the kidney function.
}
\label{flowchart}
\end{figure*}

Our methodological framework (Figure~\ref{flowchart}) consists of DMD, thresholding and selection of kidney region using connected component (blob) analysis.

In our previous studies~\cite{Tirunagari2017}, we presented a novel automated, registration-free movement correction approach
based on windowed and reconstruction variants of Dynamic Mode Decomposition (WR-DMD) to suppress unwanted complex organ motion in DCE-MRI image sequences caused due to respiration.

\subsection{Dynamic Mode Decomposition (DMD)}
\label{meth:DMD}

In a dynamic sequence of $N$ images $\textbf{P}$, let $p_r$ be the $r^{th}$ image whose size is $m\times n$. This image $p_r$ is converted to $mn\times1$ column vector, resulting in the construction of a data matrix $\textbf{P}$ of size  $mn \times N$ for $N$ images in the DCE-MRI data.

\begin{equation}\label{eq:display13} 
\textbf{P} =   [p_1, p_2, p_3, \cdots, p_N] = \begin{pmatrix}
p^{1}_1 & p^{1}_2 & ... & p^{1}_N \\
\vdots & \vdots  &  \vdots & \vdots  \\  
p^{mn}_1 & p^{mn}_2 & ... & p^{mn}_N \\
\end{pmatrix}.
\end{equation}

The images in the DCE-MRI data are collected over regularly spaced time intervals and hence each pair of consecutive images are linearly correlated. It can be justified that a linear mapping A exists between them forming a span of krylov subspace~\cite{krylov1931numerical,saad1981krylov,ruhe1984rational}: 
\begin{equation}
\textbf{P} = [p_1, Ap_1, A^2p_1, A^3p_1, \cdots, A^{N-1}p_{1}]. 
\end{equation}
The krylov subspace can then be represented using two matrices $P_2$ and $P_1$ where $P_2 \equiv [p_2, p_3, \cdots, p_N]$ and \\ $P_1 \equiv [p_1, p_2, \cdots, p_{N-1}]$. 
\begin{equation}
\label{Ap_1}
    P_2 = AP_1.
\end{equation}

The mapping matrix $A$ is responsible for capturing the dynamics or the fluctuating intensities within the image sequence. 
The sizes of the matrices $P_2$ and $P_1$ are both $mn \times N-1$ each. Therefore, the size of unknown matrix $A$ would be $mn \times mn$. Unfortunately, solving for A is computationally very expensive due to it size. For instance, if an image has a size of $240 \times 320$ i.e., $m=240$ and $n=320$, the size of $A$ is then $76800 \times 76800$.

Since solving $A$ is computationally expensive for large image dimensions, we need an alternative solution. Our assumption that the images form a Krylov  span, allows us to introduce $H$ from the standard Arnoldi iteration~\cite{trefethen1997numerical}).

\begin{equation}
\label{eq:ar1}
AP_1 \approx P_1H.
\end{equation}

Here, $H$ a companion matrix also known as a shifting matrix that simply shifts images $1$ through $N-1$ and approximates the last frame $N$ by linearly combining the previous $N-1$ images, i.e., $P_N = c_0p_1 + ... + c_Np_{N-1} =[p_1, p_2, p_3, \cdots, p_{N-1}]c$. $H$ requires the storage of $N-1 \times N-1$ data matrix is significantly smaller than $A$ in dimensions.

\begin{equation}
H = \left( \begin{array}{ccccc}
 0 & 0      & \ldots & 0       & c_0 \\
 1 & 0      & \ldots & 0       & c_1 \\
   & \ddots & \ddots & \vdots  & \vdots \\
   &        & 1      & 0       & c_{N-2} \\
   &        &        & 1       & c_{N-1} \end{array} \right).
\end{equation}

Thus, for the last frame $N$, where $N$ is much fewer (dimension) than the dimensionality of $A$ ($mn \times mn$), one can write $P_2$ as a linear combination of the previous vectors. Consistent with Equations~\ref{Ap_1} and \ref{eq:ar1}, we then have: 
\begin{equation}\label{eq:ar2}
P_2 \approx P_1H.
\end{equation}
From Equations~\ref{eq:ar1} and \ref{eq:ar2}, we have $ AP_1 \approx P_2 \approx P_1H.$

In~\cite{Schmid2}, the author describes a more robust solution, which is achieved by applying a singular value decomposition (SVD) on $P_1$. From Equation~\ref{eq:ar2}, SVD decomposition on $P_1$ subspace is calculated to obtain $U$, $\Sigma$ and $V^*$ matrices that are left singular vectors, singular values and right singular vectors respectively. The inversions of these matrices are then multiplied with $P_2$ subspace to obtain the The full-rank matrix $\widetilde{H}$,  determined  on  the  subspace  spanned  by  the  orthogonal basis  vectors $U$ of $P_1$. described by: 

\begin{equation}
\widetilde{H} = U^*P_2V\Sigma^{-1}.\\
\end{equation}

Here, $U^* \in \mathbb{C}$ and $V \in \mathbb{C}$ are the conjugate transpose of $U$ and $V^*$, respectively; and $\Sigma^{-1} \in \mathbb{C}^{N\times N}$ denote the inverse of the singular values $\Sigma$. After obtaining the $\widetilde{H}$ matrix, the eigenvalue analysis is performed to obtain $\omega$ eigenvectors and $\sigma$ a diagonal matrix containing the corresponding eigenvalues. 

\begin{equation}
\widetilde{H}\omega = \sigma\omega
\end{equation}

It is known that the eigenvalues of $\widetilde{H}$ approximate some of the eigenvalues of the full system $A$. The associated eigenvectors of $H$ provide the coefficients for the linear combination that is necessary to express the dynamics within the image sequence basis. The dynamic modes $\Psi$ are thus calculated as follows:
\begin{equation}
\label{dmv}
\Psi = P_2V\Sigma^{-1}\omega
\end{equation}

Since the eigenvalues associated with $\widetilde{H}$ are complex, it is not possible to establish the order of the dynamic modes $\Psi$ directly. Nevertheless, we calculate the absolute value for the phase-angles associated with the complex eigenvalues and the modes with unique phase-angles are selected. Doing this will remove one of the conjugate pairs in the dynamic modes which have same phase-angles but with different signs and capture similar information~\cite{sayadi2013dynamic}. After discarding one of the conjugate pairs, the dynamic modes are then sorted in the ascending order of their phase-angles.

\section{Dataset}
\label{data}
In this section we briefly describe our synthetically generated data as well as the DCE-MRI data. 

\begin{figure}[!h]
\centering
\begin{tabular}{cc}
\includegraphics[scale=0.3]{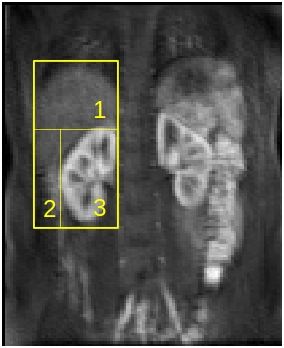} & \includegraphics[width=0.25\textwidth]{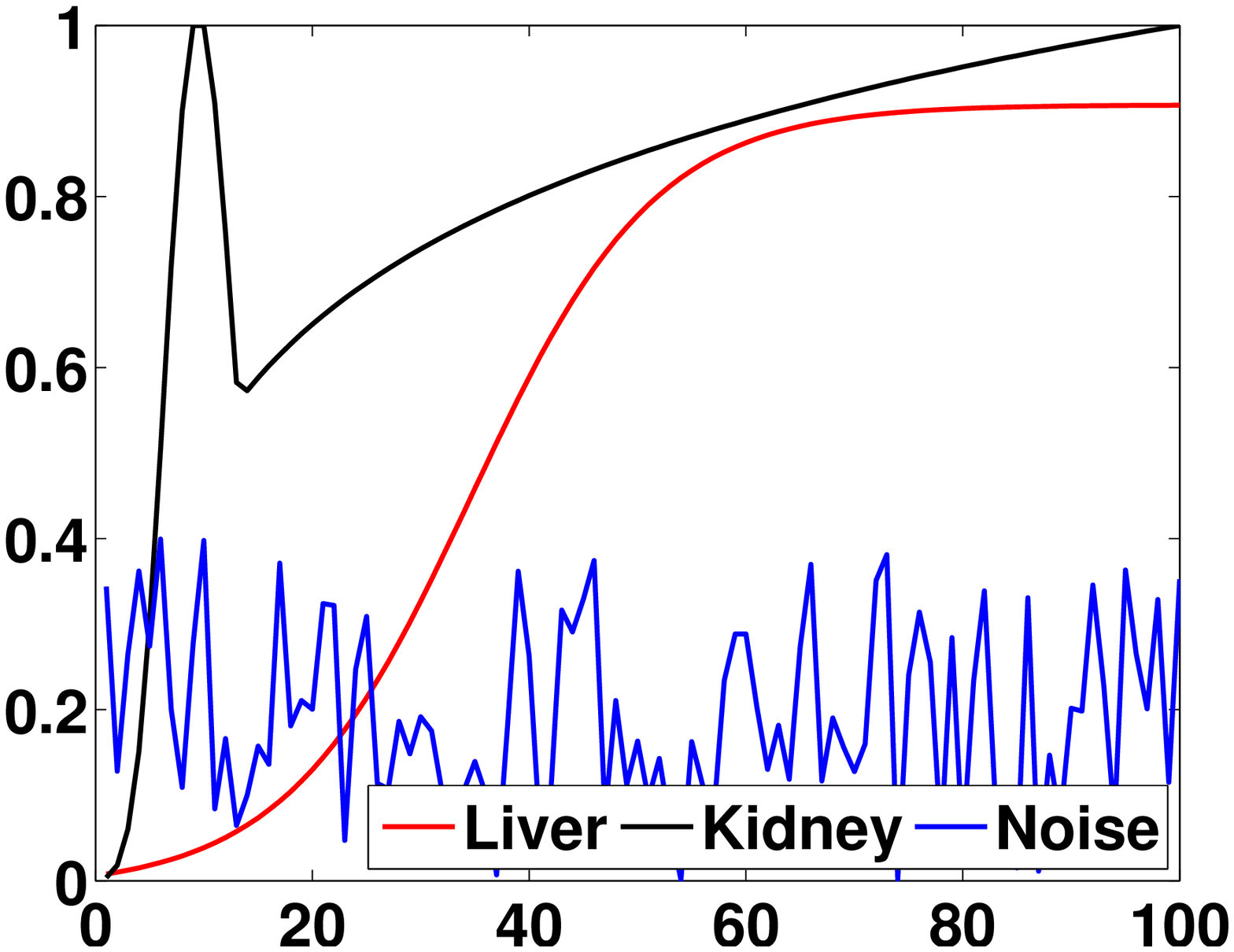} \\
(a) & (b) \\
\includegraphics[width=0.22\textwidth]{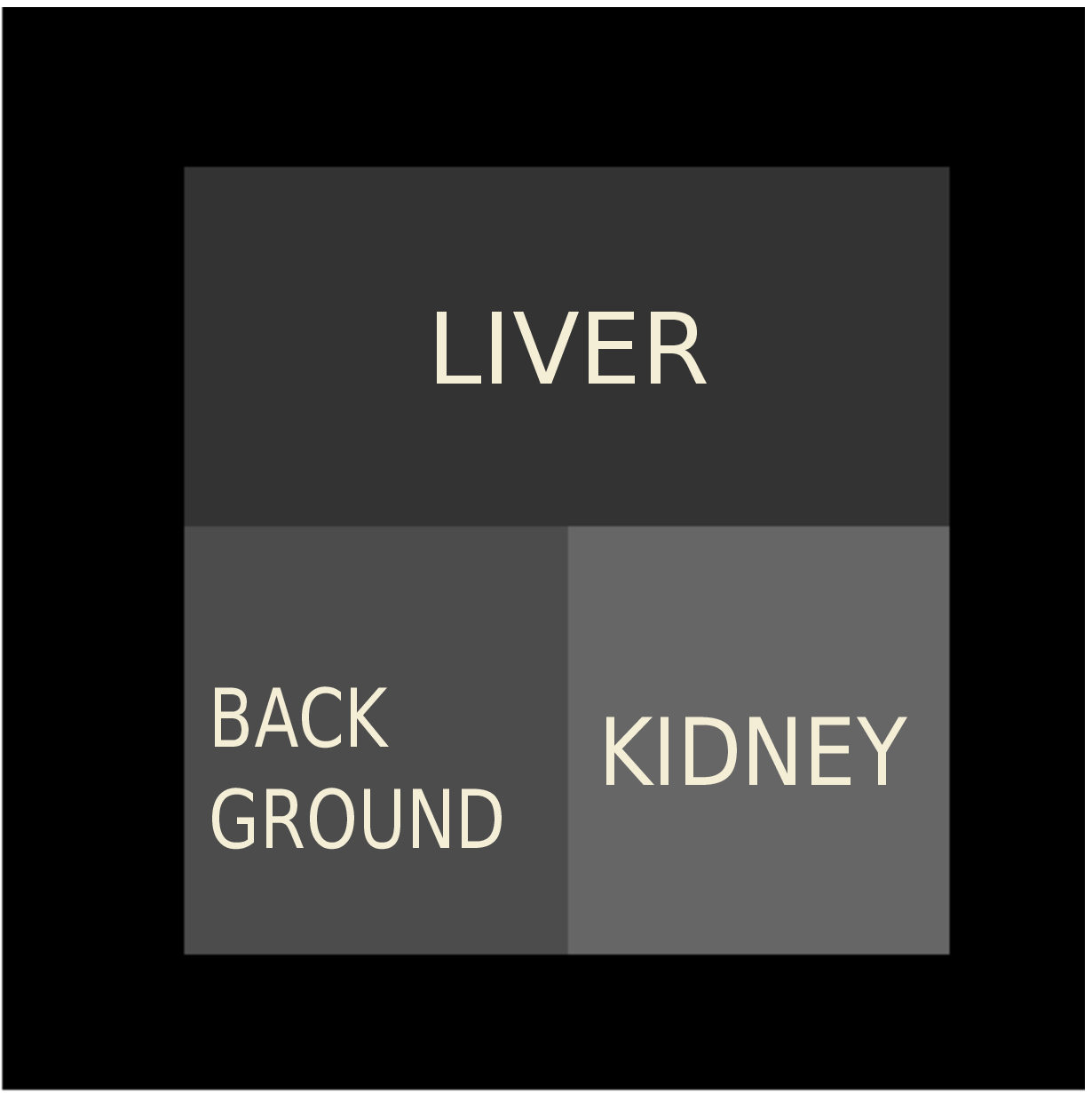} & \includegraphics[width=0.22\textwidth]{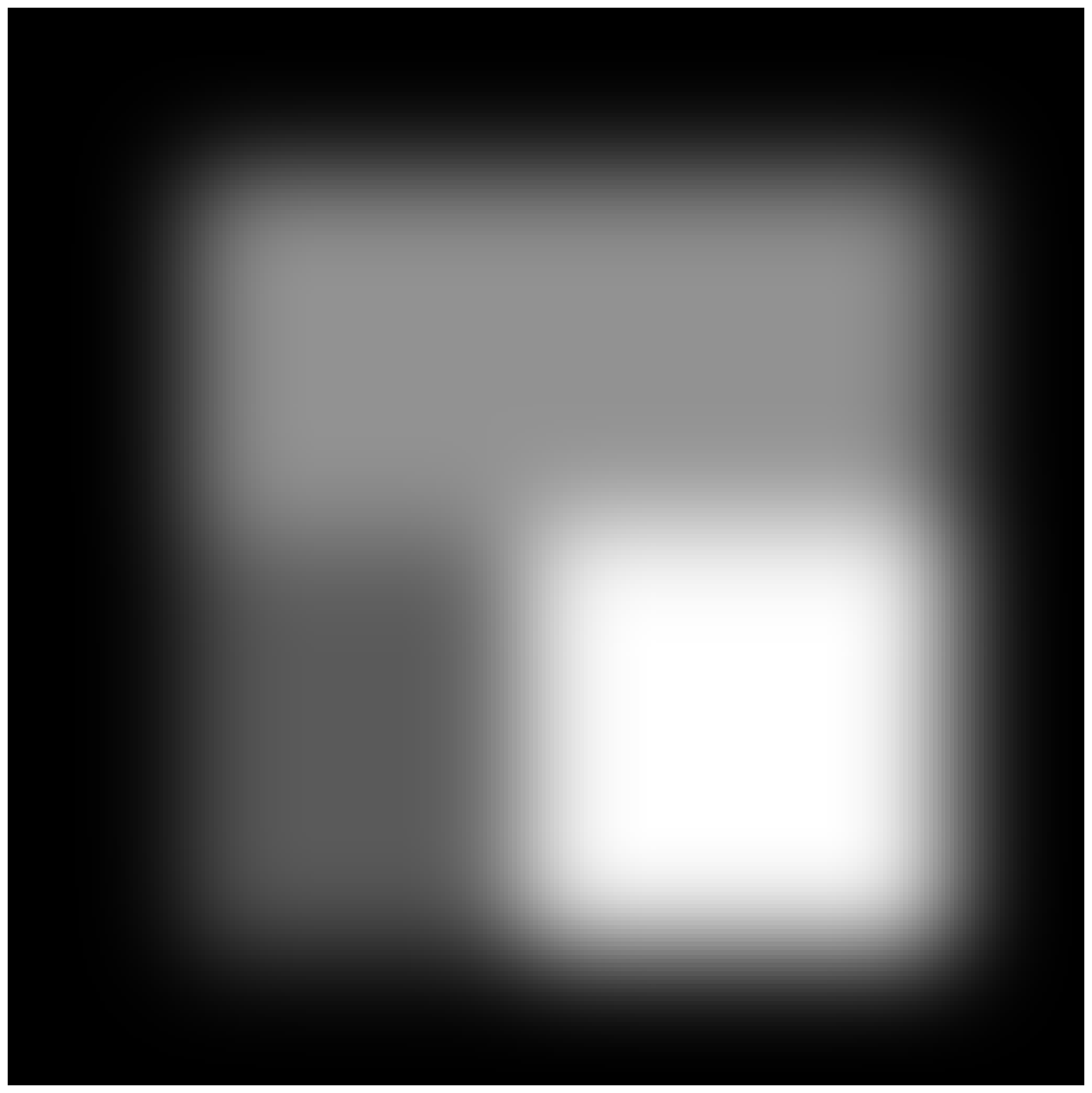}\\
(c) & (d)\\
\end{tabular}
\caption{(a) A random image from a DCE-MRI sequence showing the regions of liver (1), background (2) and kidney (3) regions. (b) Graph showing the time-intensity evolution of idealised kidney, liver and background functions for synthetically generated data. (c) Simulated image representing kidney, liver and background functions corresponding to the DCE-MRI image as shown in (a). (d) Gaussian PSF Convolved idealised representation of the image showing the effects of PVC i.e., by mixing up of regions especially near the borders.}
\label{fulspec-2}
\end{figure}

\subsection{Synthetic data}

\begin{figure*}
\centering
\begin{tabular}{ccccc}
\includegraphics[scale=0.35]{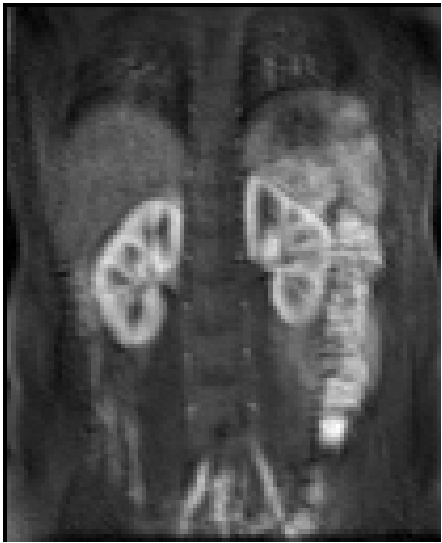} & \includegraphics[scale=0.35]{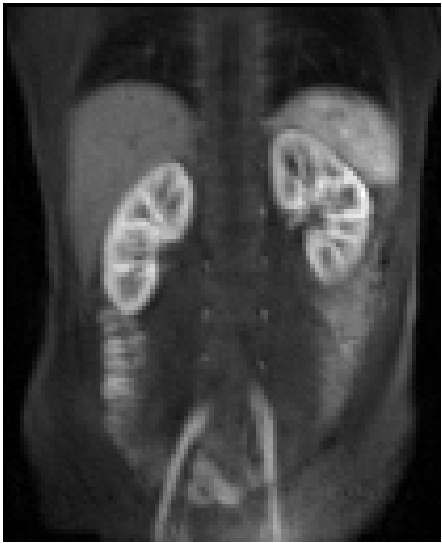} & \includegraphics[scale=0.35]{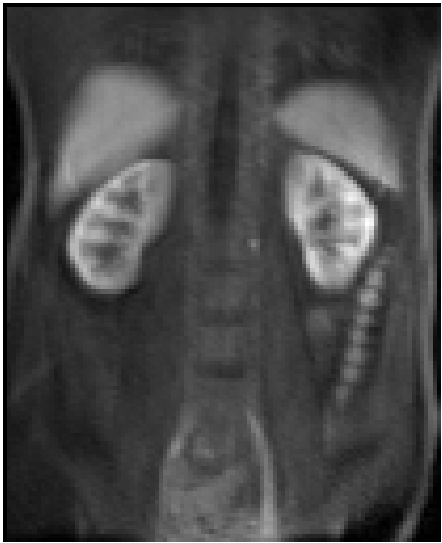} & \includegraphics[scale=0.35]{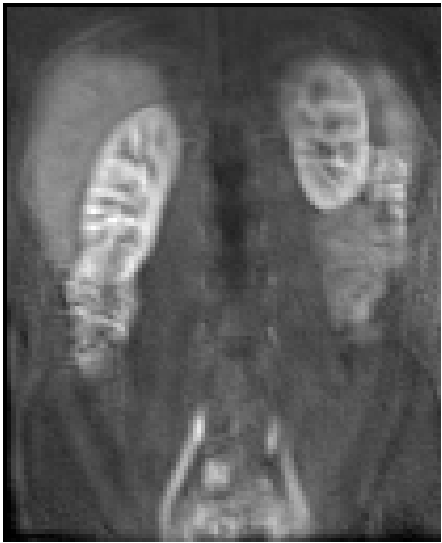} & \includegraphics[scale=0.35]{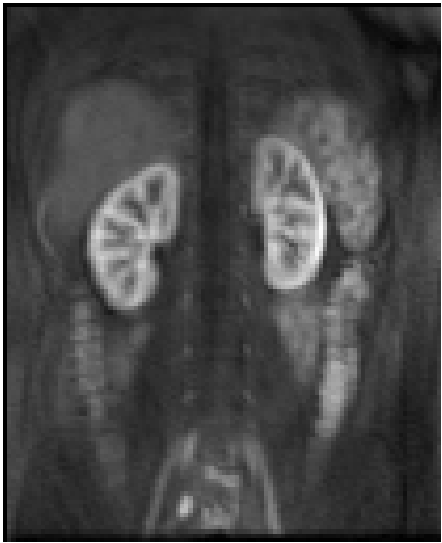}\\
(1) & (2) & (3) & (4) &(5) \\
\includegraphics[scale=0.35]{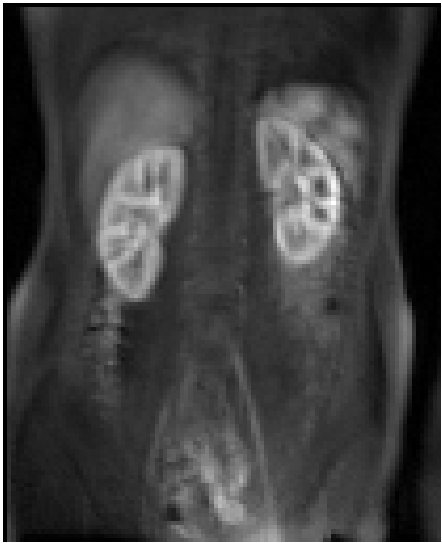} & \includegraphics[scale=0.35]{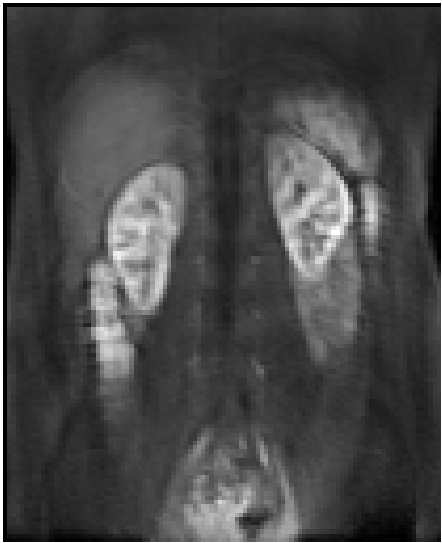} & \includegraphics[scale=0.35]{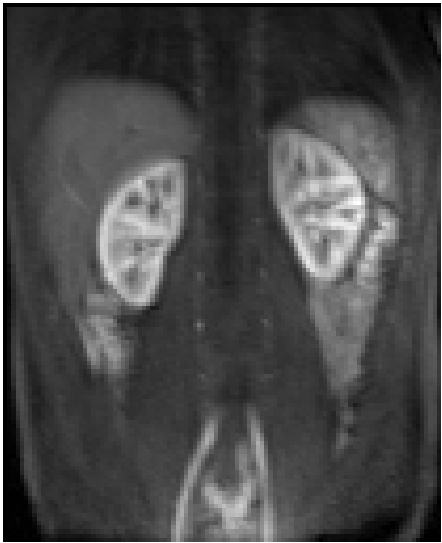} & \includegraphics[scale=0.35]{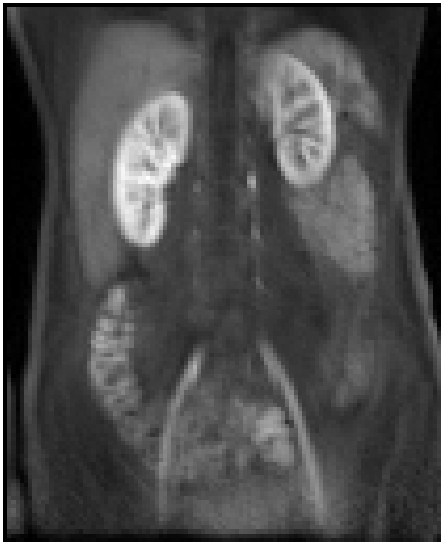} & \includegraphics[scale=0.35]{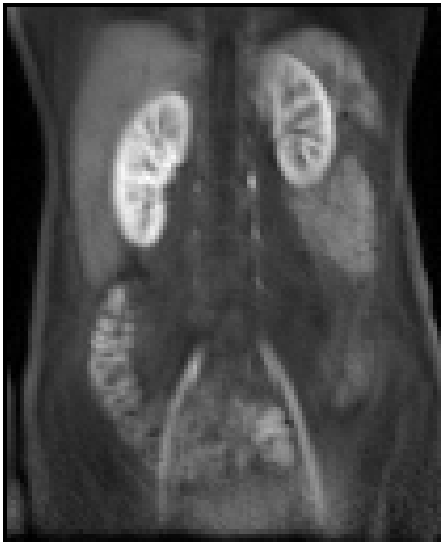}\\
(6) & (7) & (8) & (9) &(10) \\
\end{tabular}
\caption{Exemplars of Dynamic MR images from 10 healthy volunteers' kidney slice produced by DCE-MRI sequence considered as 10 different datasets in this study. The images here show the central kidney slice at time $120s$ aortic peak enhancement after the contrast agent is injected. The yellow boundary on the kidneys are a results of manual delineation from a human expert. The mean intensity values are calculated in this region across the time producing time-intensity plots.}
\label{dataset-DCE}
\end{figure*}

In order to validate our experiments with the ground-truth available, we artificially generated synthetic data corresponding to the DCE-MRI events as that shown in Figure~\ref{fulspec-2}(c). That is, as the liver region denoted by $1$ in DCE-MRI image (Figure~\ref{fulspec-2}(a)) covers both background region ($2$) and kidney region ($3$), similarly the synthetically generated images follow the same structure as shown in Figure~\ref{fulspec-2}(c).  

A series of $100$ images are then produced representing  kidney, liver and background regions/functions as labelled in Figure~\ref{fulspec-2}(c) with respect to their temporal evolution shown in Figure~\ref{fulspec-2}(b). The kidney region's temporal evolution is thus modelled as a combination of Poisson and log distributions. The liver region is modelled as a sigmoid distribution and the background region as a random normal distribution. The generated regions are then convoluted using a Gaussian kernel operator (variance = 22 and block size of $40\times40$). This operator thus simulates the point spread function's (PSF's) pixel-level mixing of the kidney regions as shown in Figure~\ref{fulspec-2}(d). 

The time-intensity plot of convolved kidney region thus shows the influence of the liver (sigmoid) and background functions as shown in Figure~\ref{res:fig7}. Similarly the time-intensity plots of liver are influenced by background and kidney regions and vice versa.

\subsection{DCE-MRI data}
The functional kidney DCE-MRI datasets used in this study have been obtained from ten healthy volunteers as shown in Figure~\ref{dataset-DCE}. These datasets are acquired after injection of $0.05$ $mmol/kg$ of Gd-DTPA (Magnevist) contrast agent, on a $1.5T$ Siemens Avanto MRI scanner, using a 32 channel body phased array coil. The MRI acquisition sequence consisted of a 3D spoilt gradient echo sequence utilising an Echo time, TE of 0.6ms, Repetition time, TR of 1.6ms, and a flip angle, FA = 17 degrees with a  temporal resolution of $1.5s$ collected for $180s$. The acquired DCE-MRI datasets cover the abdominal region, enclosing left and right kidneys and abdominal aorta.

\section{Experiments \& Results} 
\label{exp_res}
In this section, we present our experimental procedure as well as the results.  

\subsection{Evaluating the DMD framework over synthetically generated data} 
\begin{figure}[!t]
\centering
\setlength{\tabcolsep}{2pt}
\begin{tabular}{|c|c|c|c|}
\hline
\rotatebox{90}{\makebox[1.4cm][c]{\emph{\tiny Kidney}}} &
\includegraphics[scale=0.2]{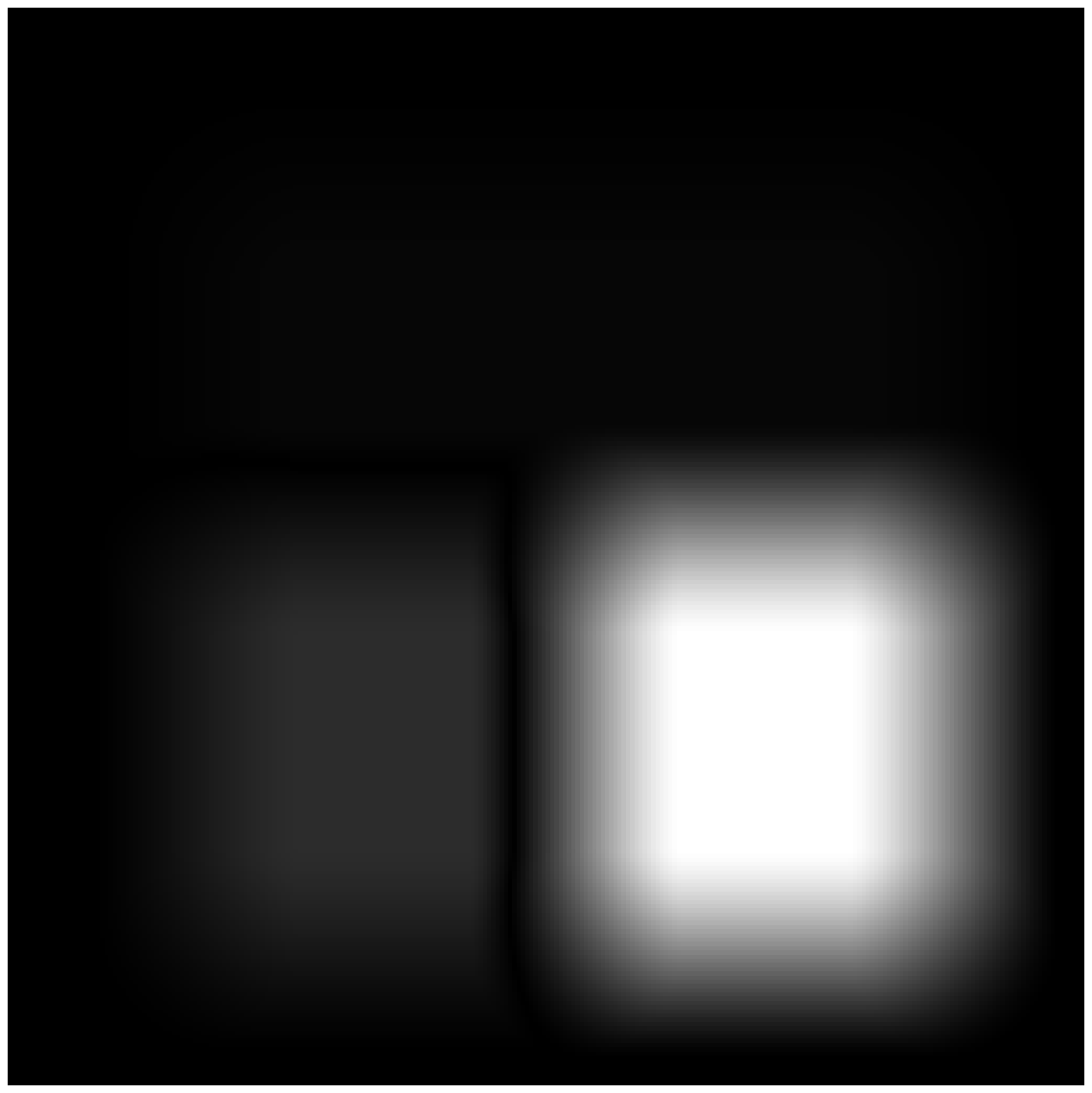} & \includegraphics[scale=0.2]{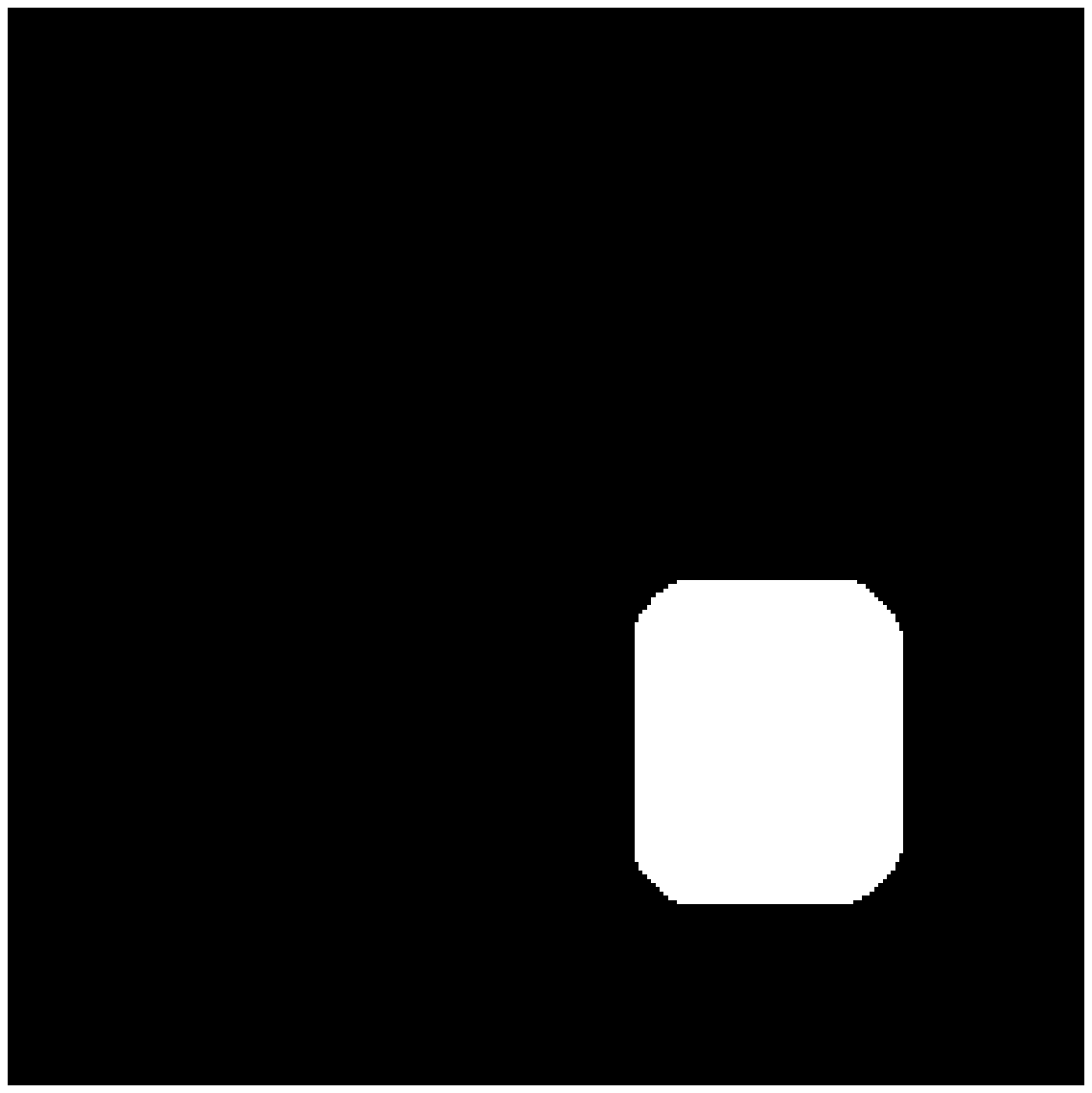} &  \includegraphics[scale=0.16]{pictures/simulated/DMDPSF-Kidney-25.eps} \\\hline
\rotatebox{90}{\makebox[1.4cm][c]{\emph{\tiny Liver}}} &
\includegraphics[scale=0.2]{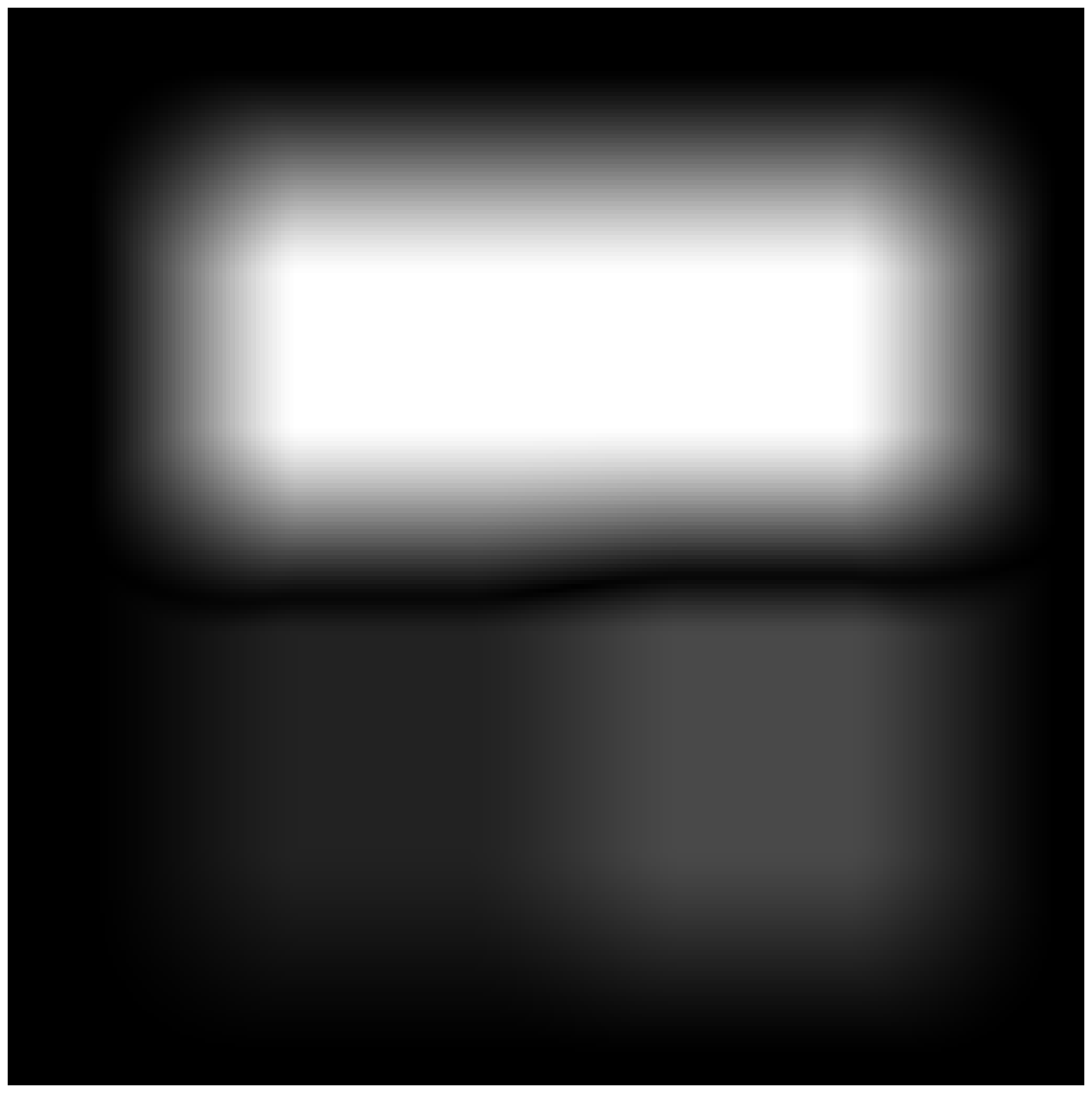} &  \includegraphics[scale=0.2]{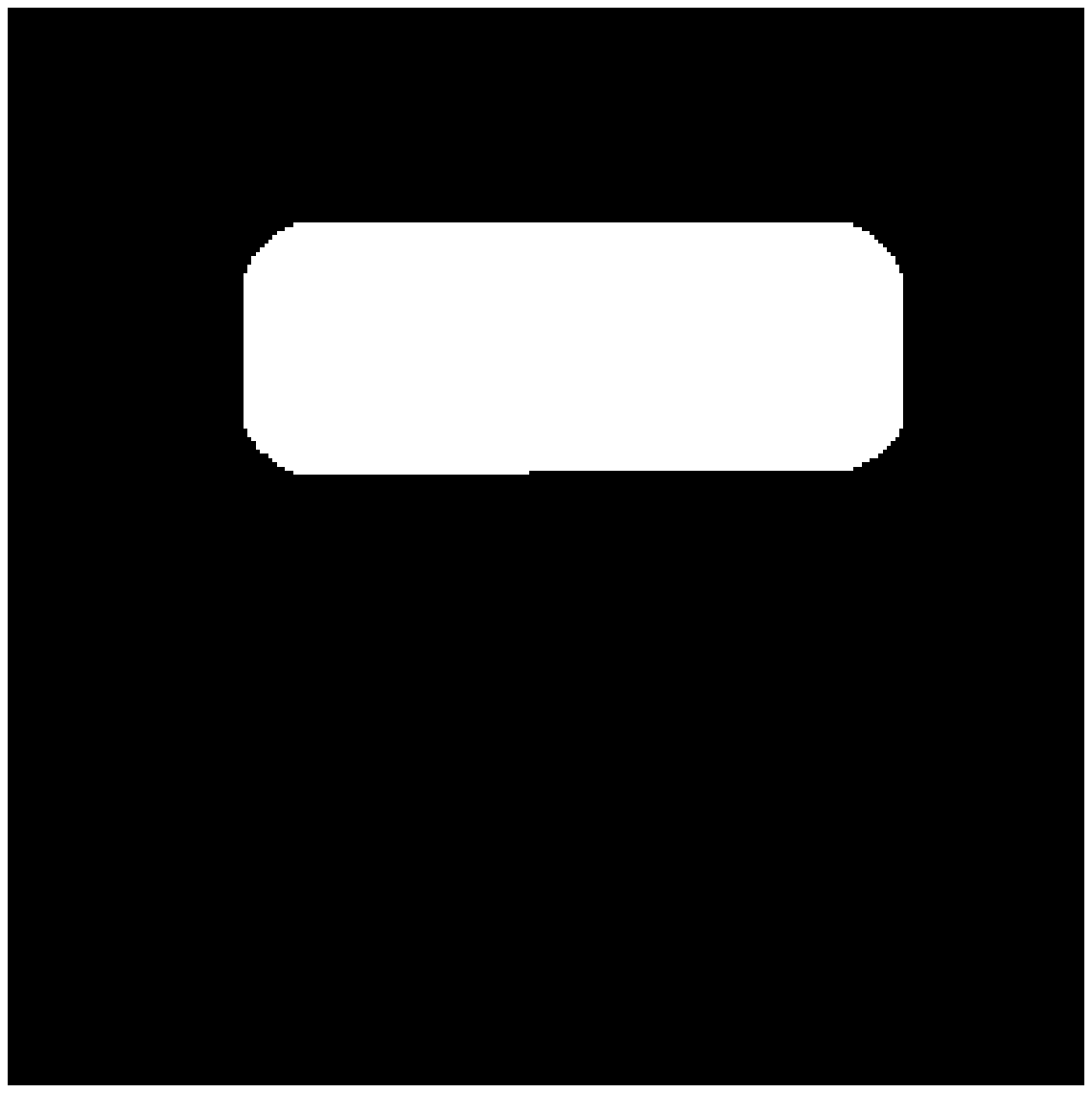} &\includegraphics[scale=0.16]{pictures/simulated/DMDPSF-Liver-25.eps} \\\hline
\rotatebox{90}{\makebox[1.4cm][c]{\emph{\tiny Background}}} &
\includegraphics[scale=0.2]{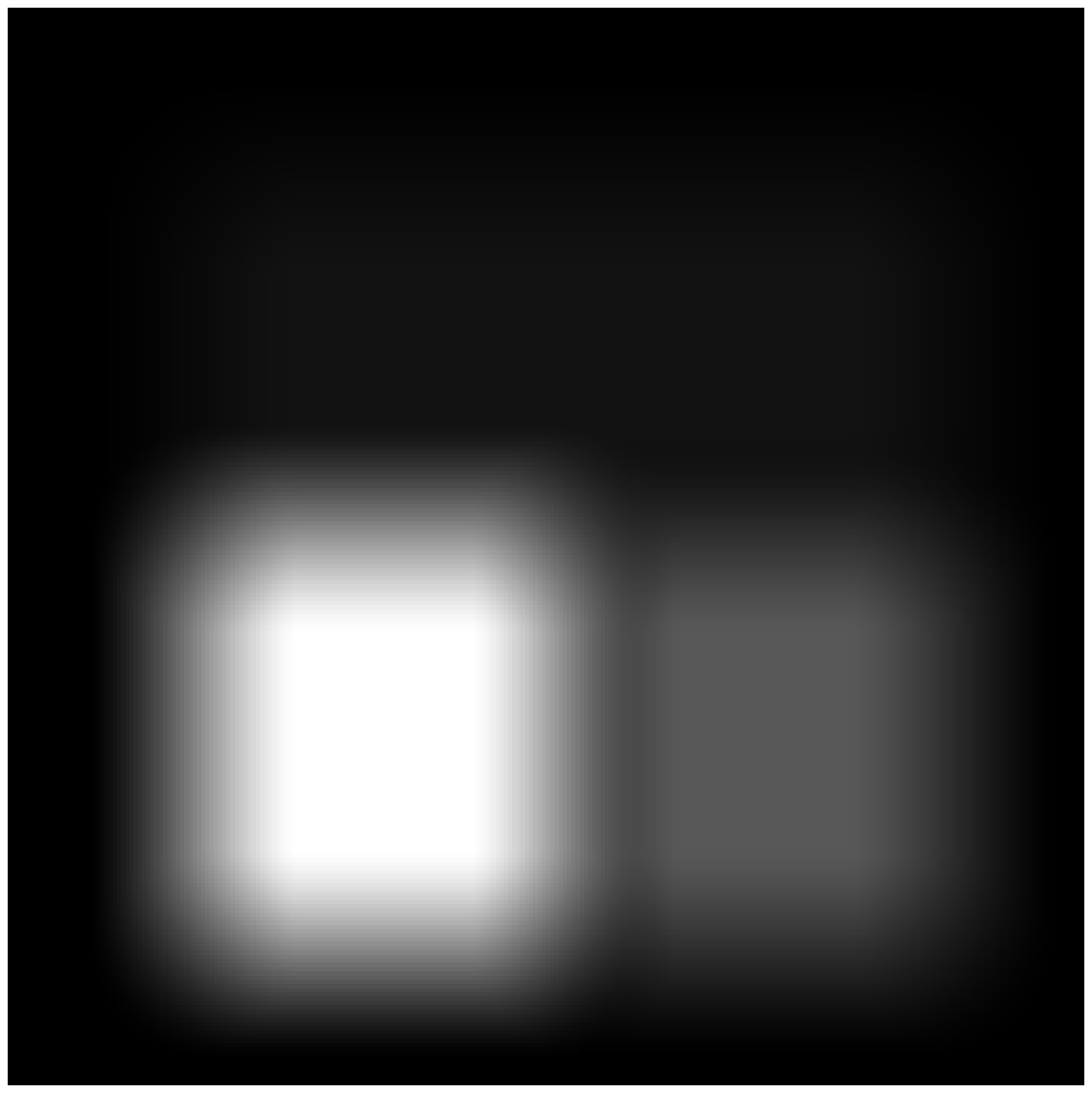} & \includegraphics[scale=0.2]{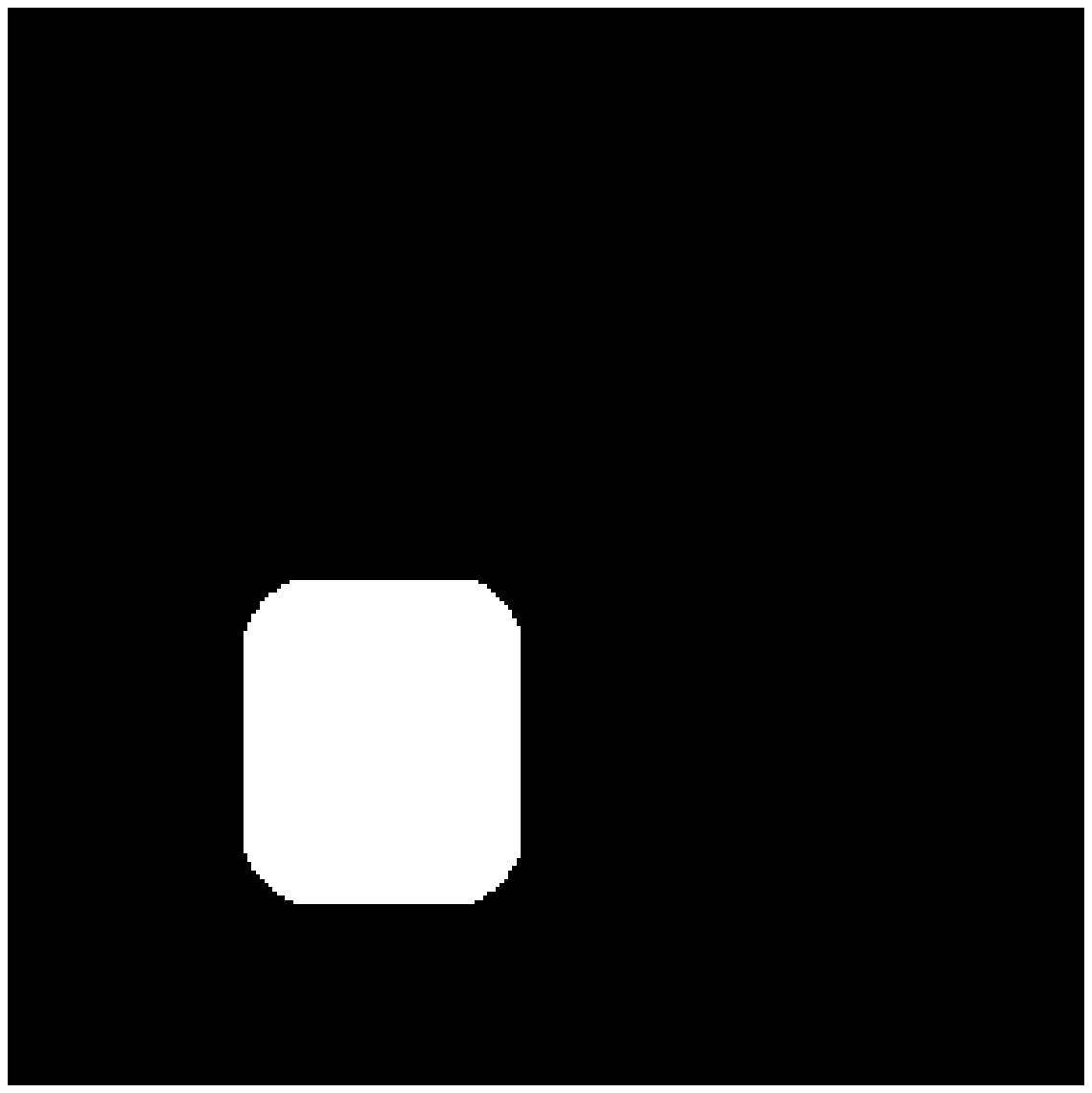} & \includegraphics[scale=0.16]{pictures/simulated/DMDPSF-background-25.eps}\\\hline
 & (a) Dynamic modes & (b) Thresholded & (c) PVC\\\hline
\end{tabular}
\caption{(a) Selection of DMD modes 1, 2 and 7. (b) Thresholding and extraction of synthetic regions. (c) Quantification of kidney, liver and background regions showing the partial volume correction (PVC).}
\label{res:fig7}
\end{figure}

Synthetic data consisting of $100$ images is given as an input to DMD algorithm which produces $99$ DMD modes (In theory, for a image sequence with $N$ images, we obtain $N-1$ DMD modes). Recall that each  dynamic mode captures a ``principal dynamics'' axes of the image sequence. Modes that show pre-dominating liver, kidney and background functions are selected, as shown in Figure~\ref{res:fig7} (a). These modes are then thresholded to obtain the segmented versions of the respective functions as shown in Figure~\ref{res:fig7} (c). These segmented versions of the dynamic modes are then projected on to the convolved image sequence and mean pixel intensities of synthetic kidney, liver and background functions are computed. The time-intensity plots are shown in Figure~\ref{res:fig7} (c). It is clear that the mixing from the other functions and are totally suppressed and the time-intensity plots produced by the proposed framework recovers the original functions.  

\subsection{Evaluating the DMD framework on DCE-MRI} 

\begin{figure*}[!t]
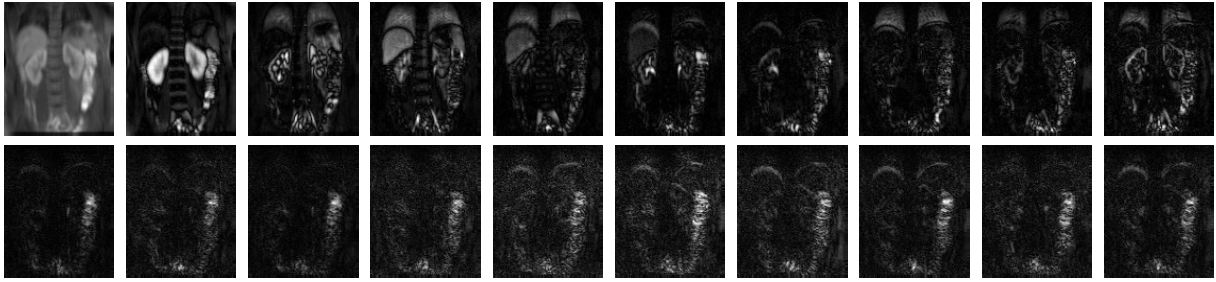

\centering
\begin{tabular}{c}
\includegraphics[scale=0.8]{images/MSDMD.eps}\\
\includegraphics[scale=0.8]{images/LSDMD.eps}\\
\end{tabular}
\caption{(Top) The top 10 ranked DMD modes on dataset-1. (Bottom) 10 images showing low ranked DMD modes on dataset-1.}
\label{MSLSDMD}
\end{figure*}

DCE-MRI image sequences are given as an input to the DMD algorithm. The output of the DMD captures large (Figure~\ref{MSLSDMD}(Top)) and small scale structures (Figure~\ref{MSLSDMD}(bottom)) in the form of modes. Dynamic mode-1 reveals the background (low-rank) model and the remaining $118$ modes capture the sparse representations. The contrast changes are captured in the top modes, particularly, mode-2 capturing kidney region and mode-3 and 4, the spleen and the liver regions respectively, as shown in Figure~\ref{MSLSDMD}(top) on the dataset-1. Noise and residuals are captured in the lower modes (Figure~\ref{MSLSDMD}(bottom)). 

\begin{figure*}[!t]
\centering
\begin{tabular}{c}
\includegraphics[scale=0.8]{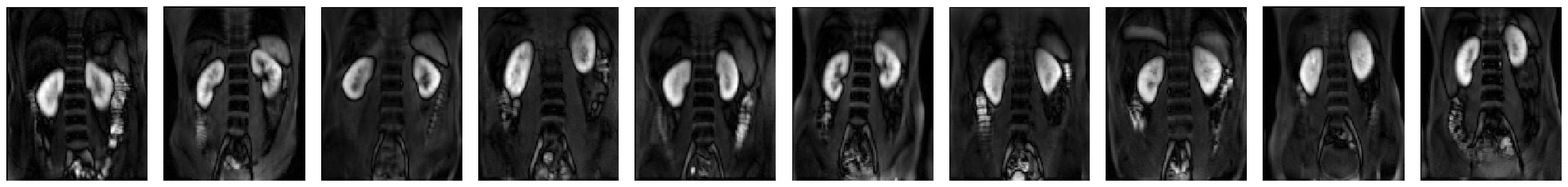} \\
(a) Dynamic mode-2 across 10 datasets of DCE-MRI data\\ 
\includegraphics[scale=0.8]{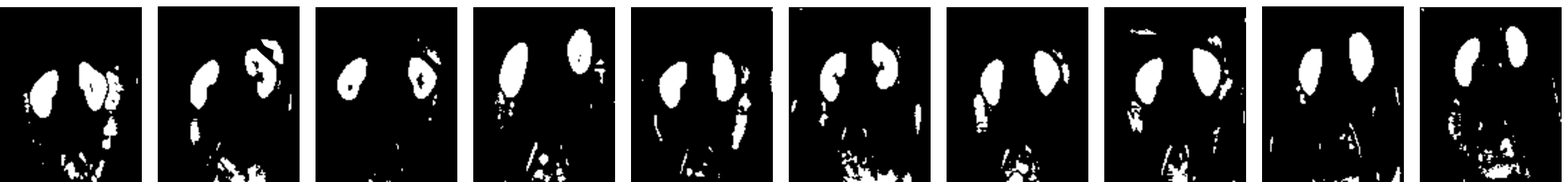}\\ 
Thresholding across 10 datasets of DCE-MRI data\\
\includegraphics[scale=0.8]{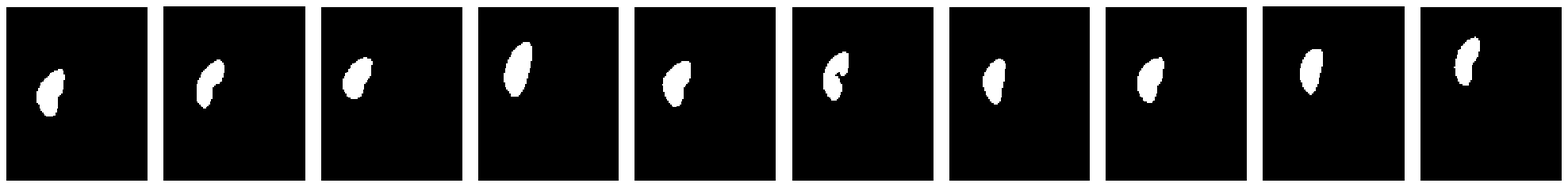} \\ 
Delineated kidney region obtained via blob analysis across 10 datasets of DCE-MRI data\\
\end{tabular}
\caption{(a) Images showing the kidney regions in Dynamic mode-2 across the 10 datasets used in this study. (b) Images showing the thresholding effect on dynamic mode-2 across the datasets. (c) Images showing the kidney template selected by the largest area of connected components. }
\label{result}
\end{figure*}

Therefore, we have selected DMD mode-2 across all the 10 datasets as shown in Figure~\ref{result} (a). These modes are then thresholded to obtain a binary of version (Figure~\ref{result} (b)). Later using connected component analysis on the left half of the image, the largest blob is selected as a kidney ROI as shown in Figure~\ref{result} (c).  The mean of intensity values inside this automatically delineated kidney ROI are calculated to produce time-intensity plots.

\subsection{Comparison with human expert} 
\begin{figure*}[h]
\setlength{\tabcolsep}{2pt}
\begin{tabular}{cccc}
\includegraphics[scale=0.25]{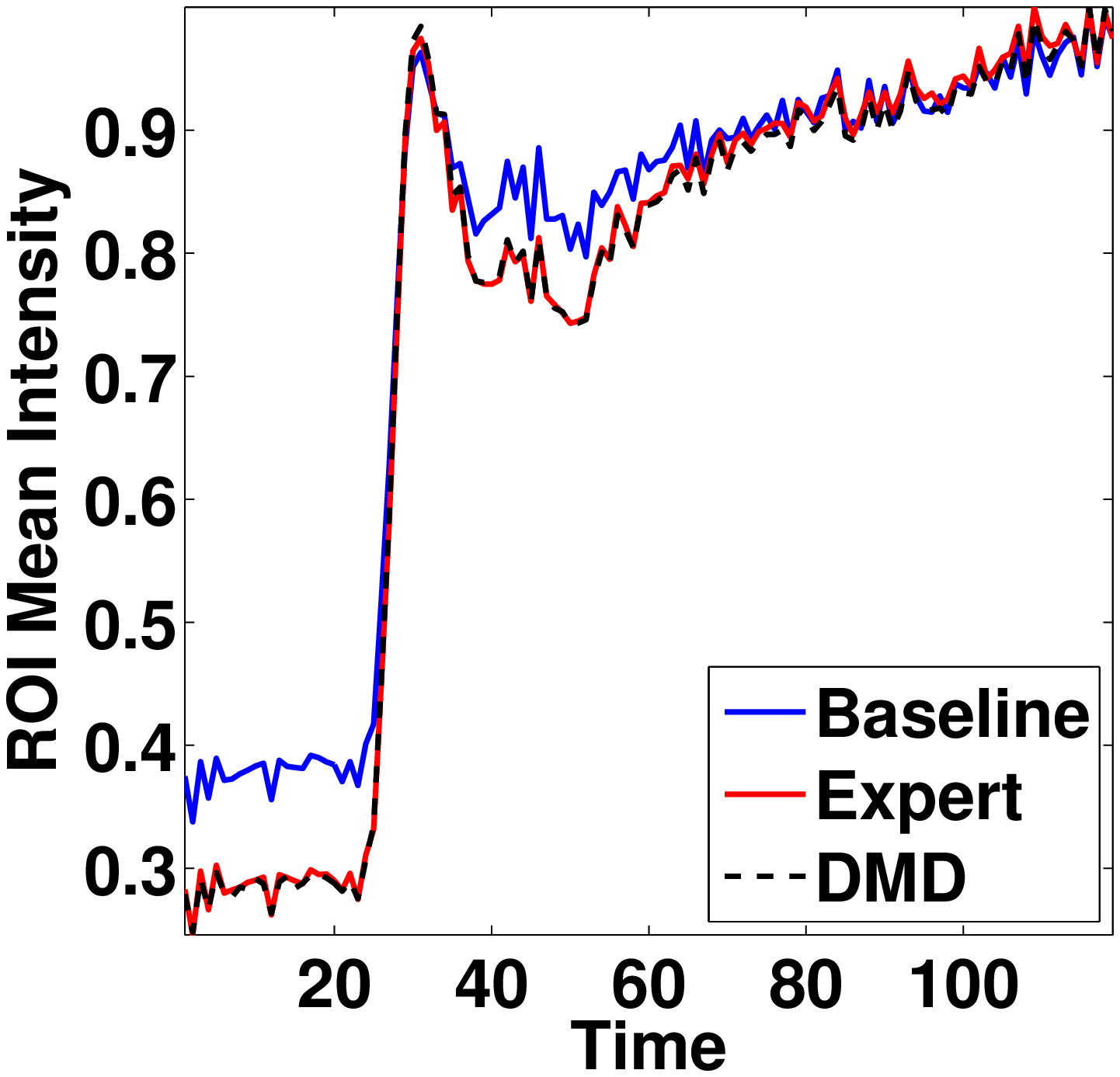} & \includegraphics[scale=0.25]{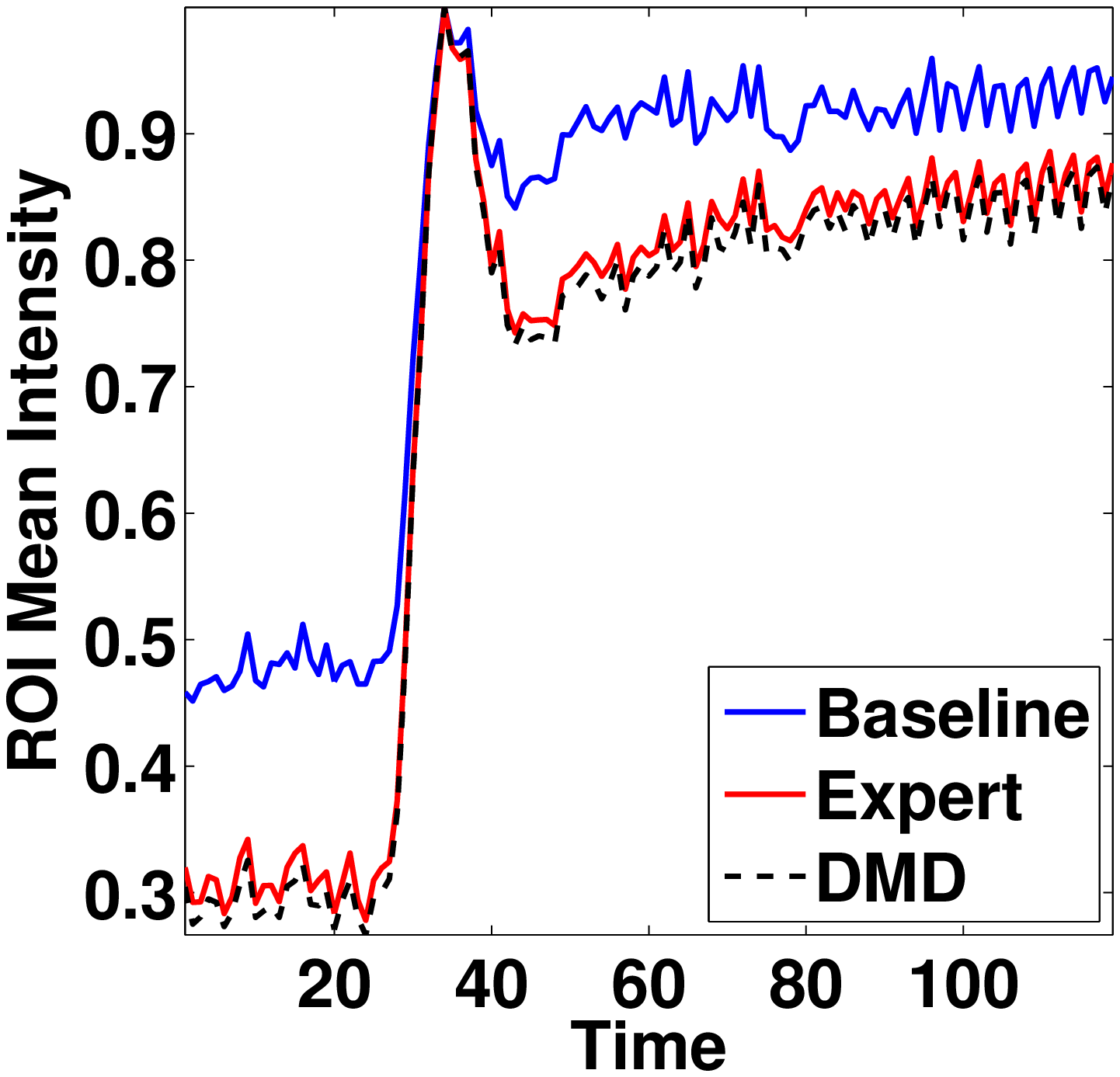} & \includegraphics[scale=0.25]{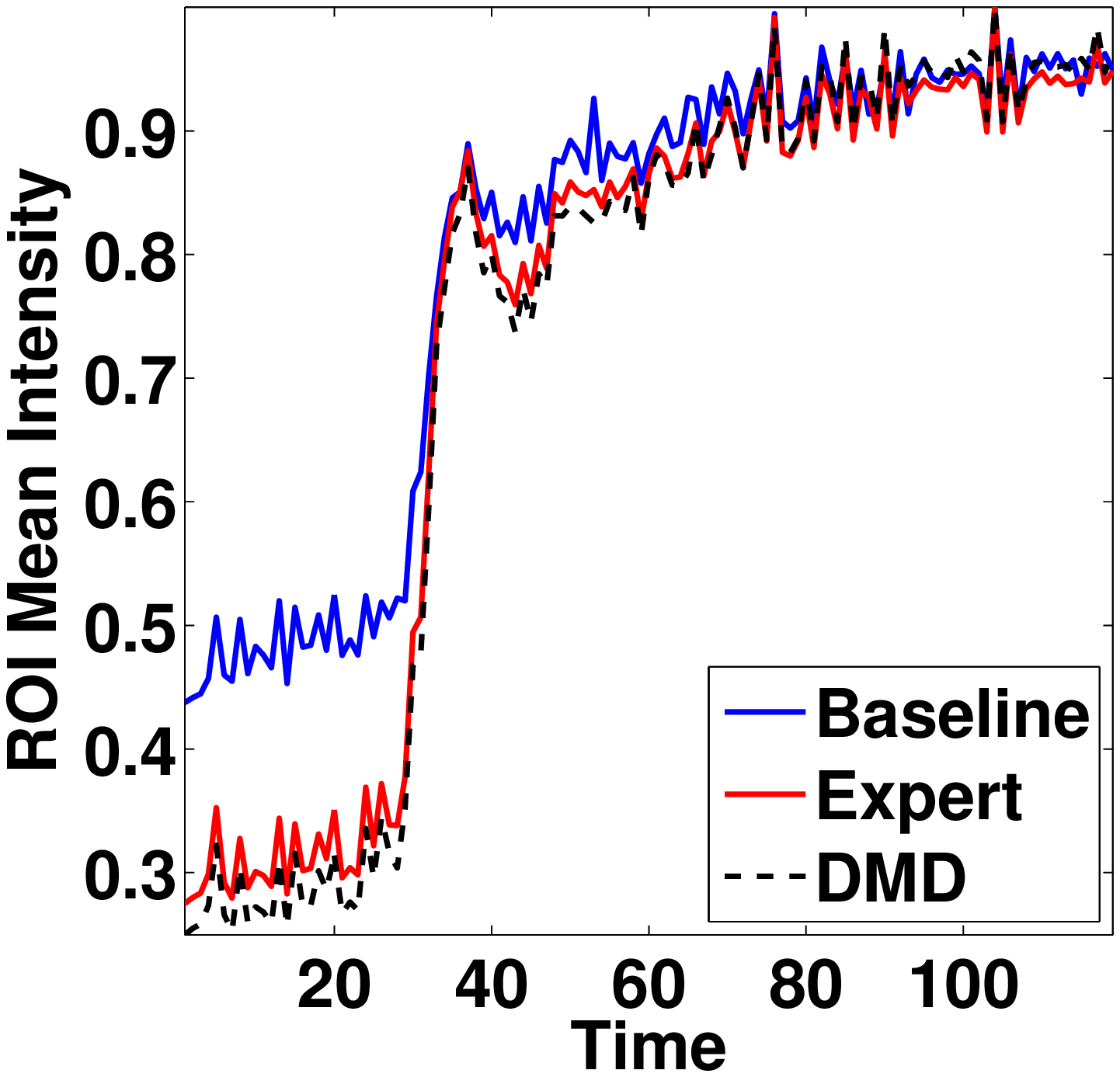}& \includegraphics[scale=0.25]{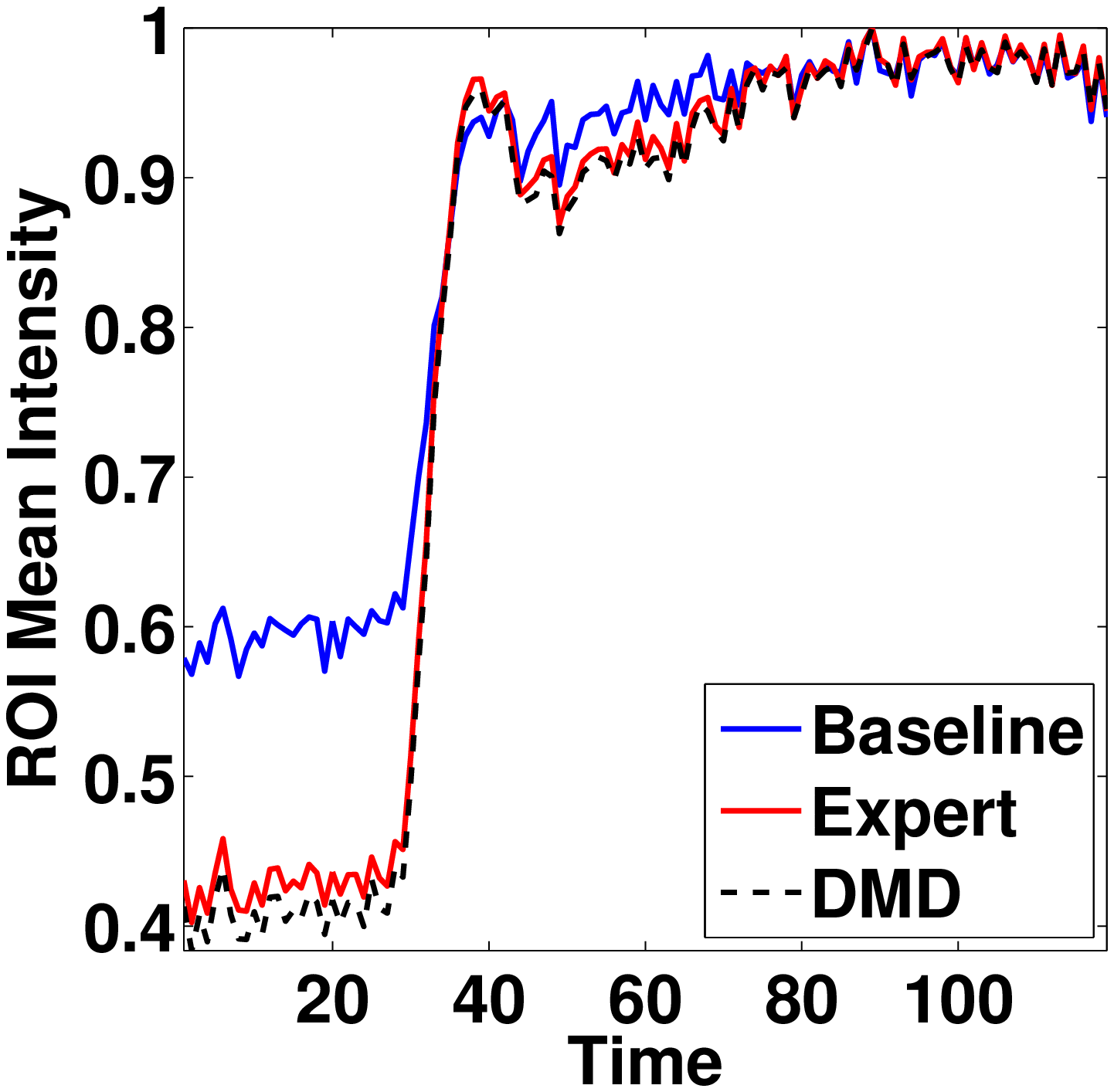} \\ 
(1) & (2) & (3) & (4) \\
\includegraphics[scale=0.25]{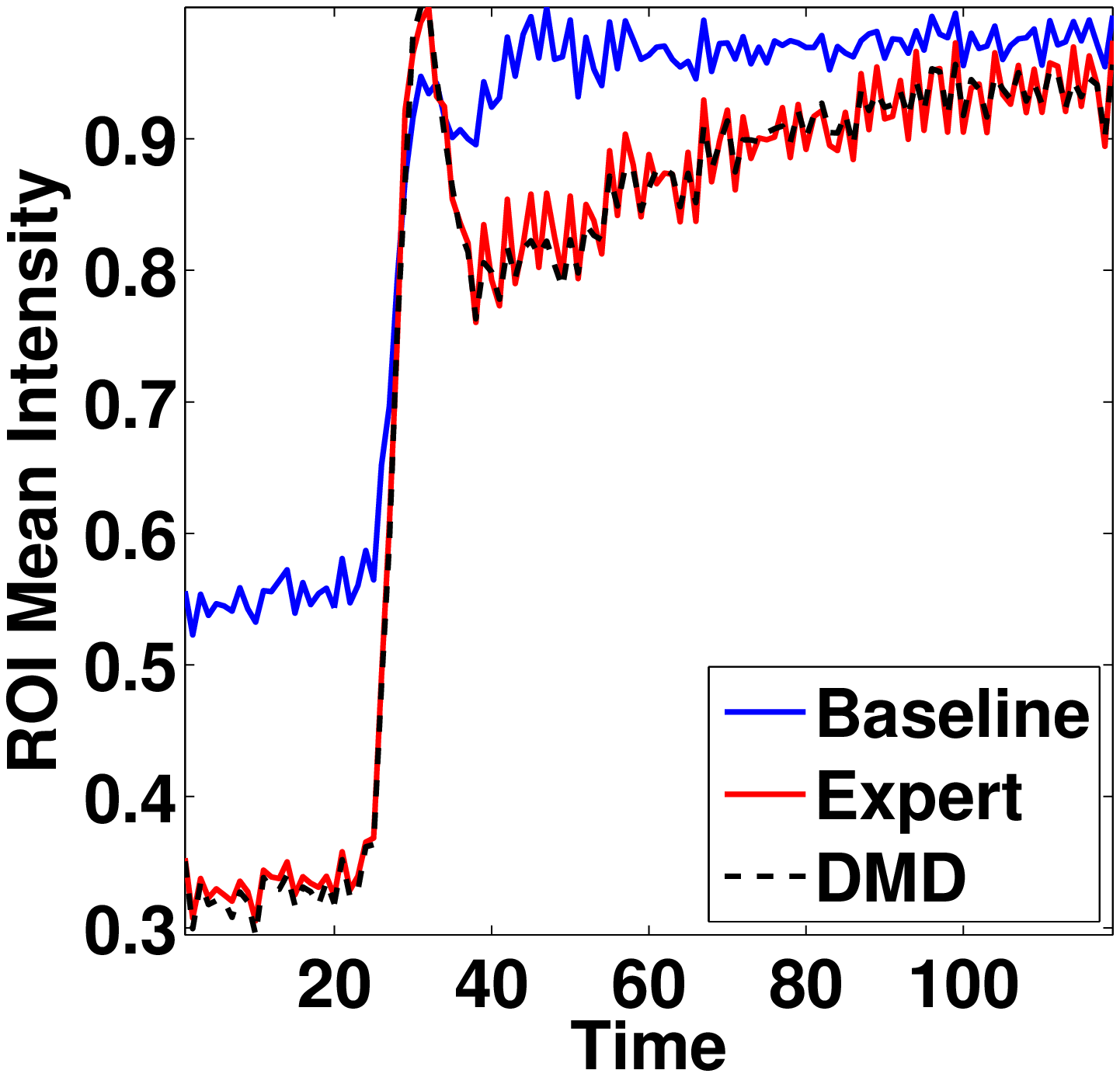} & \includegraphics[scale=0.25]{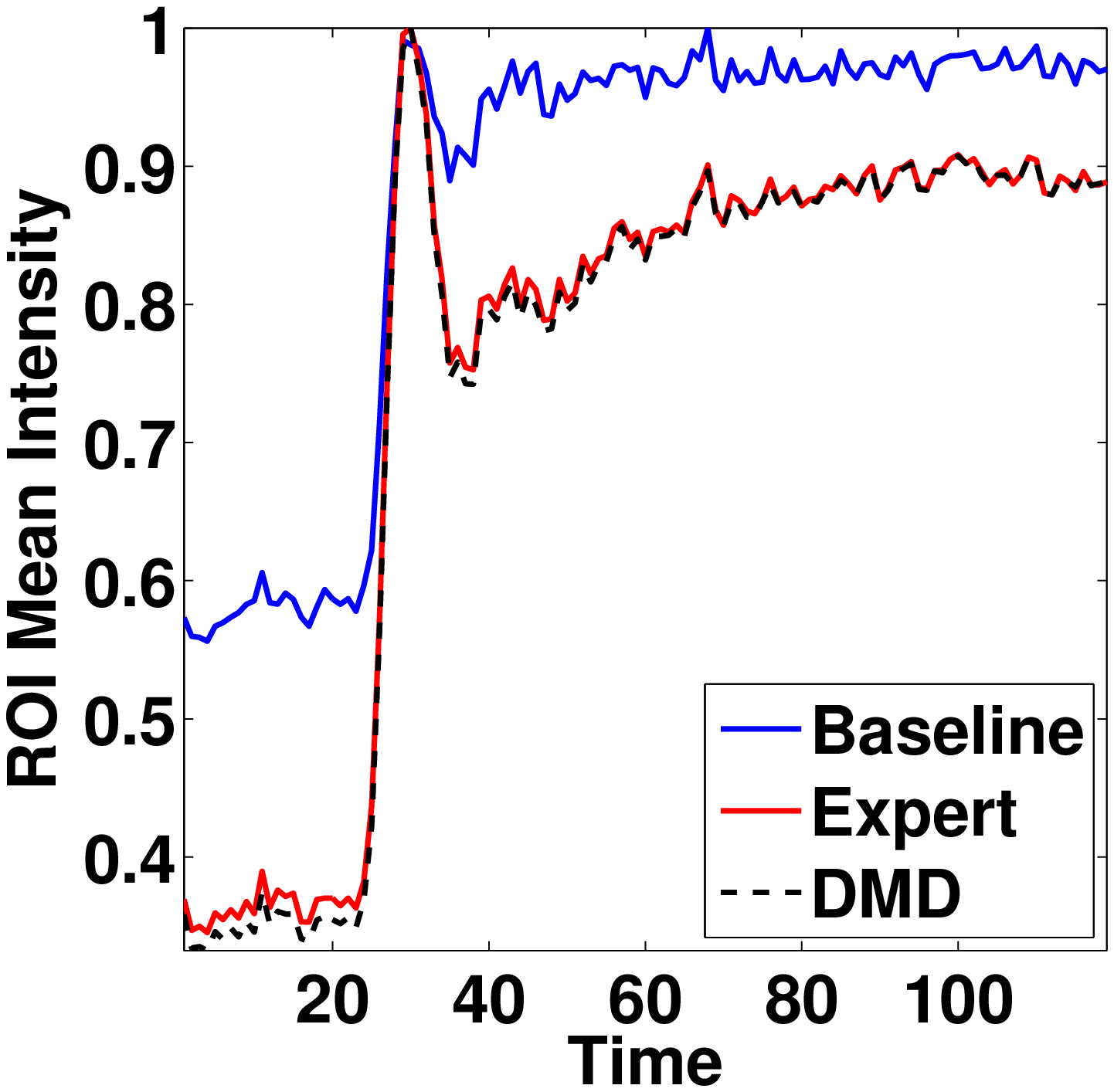} & \includegraphics[scale=0.25]{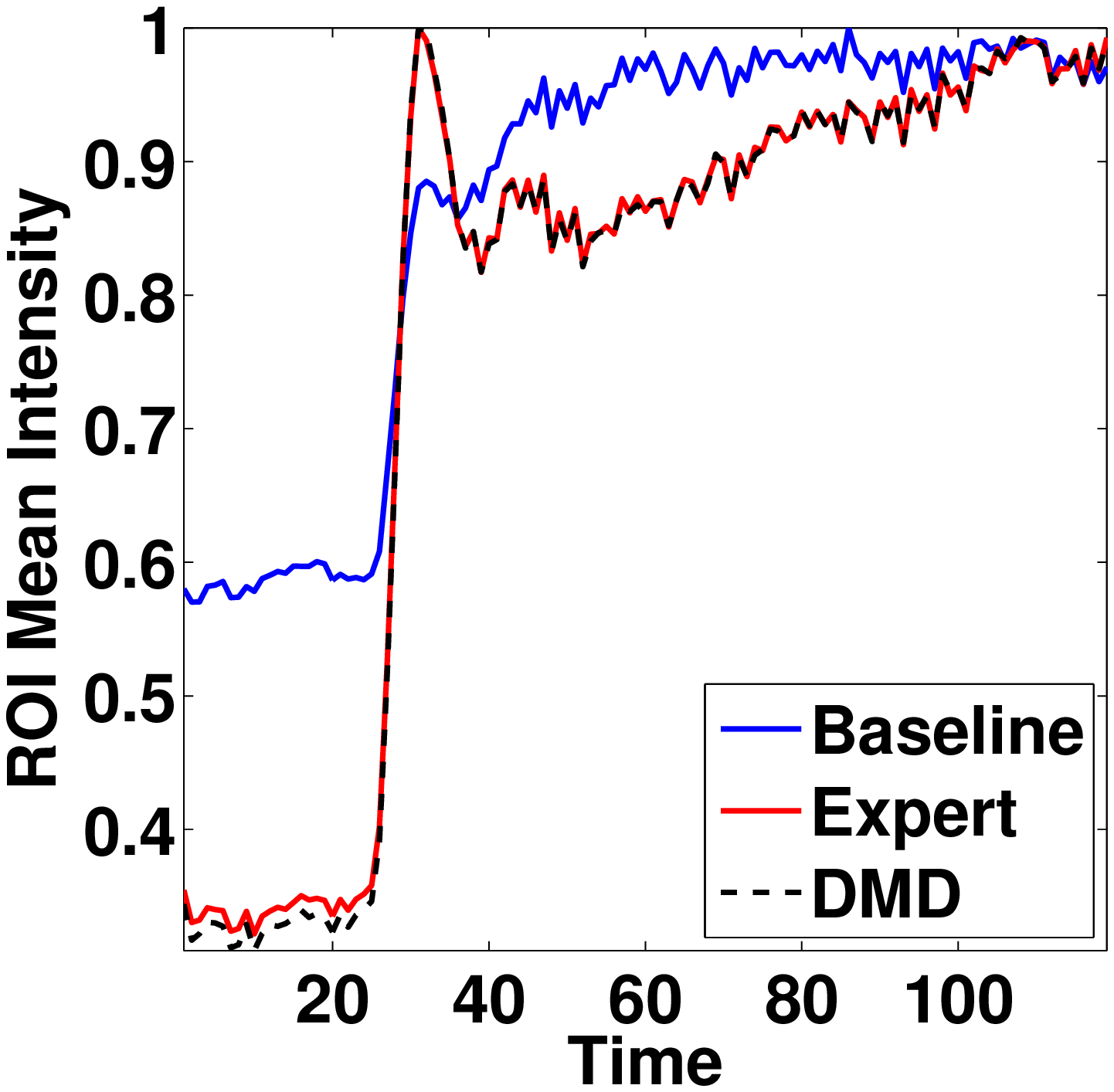} & \includegraphics[scale=0.25]{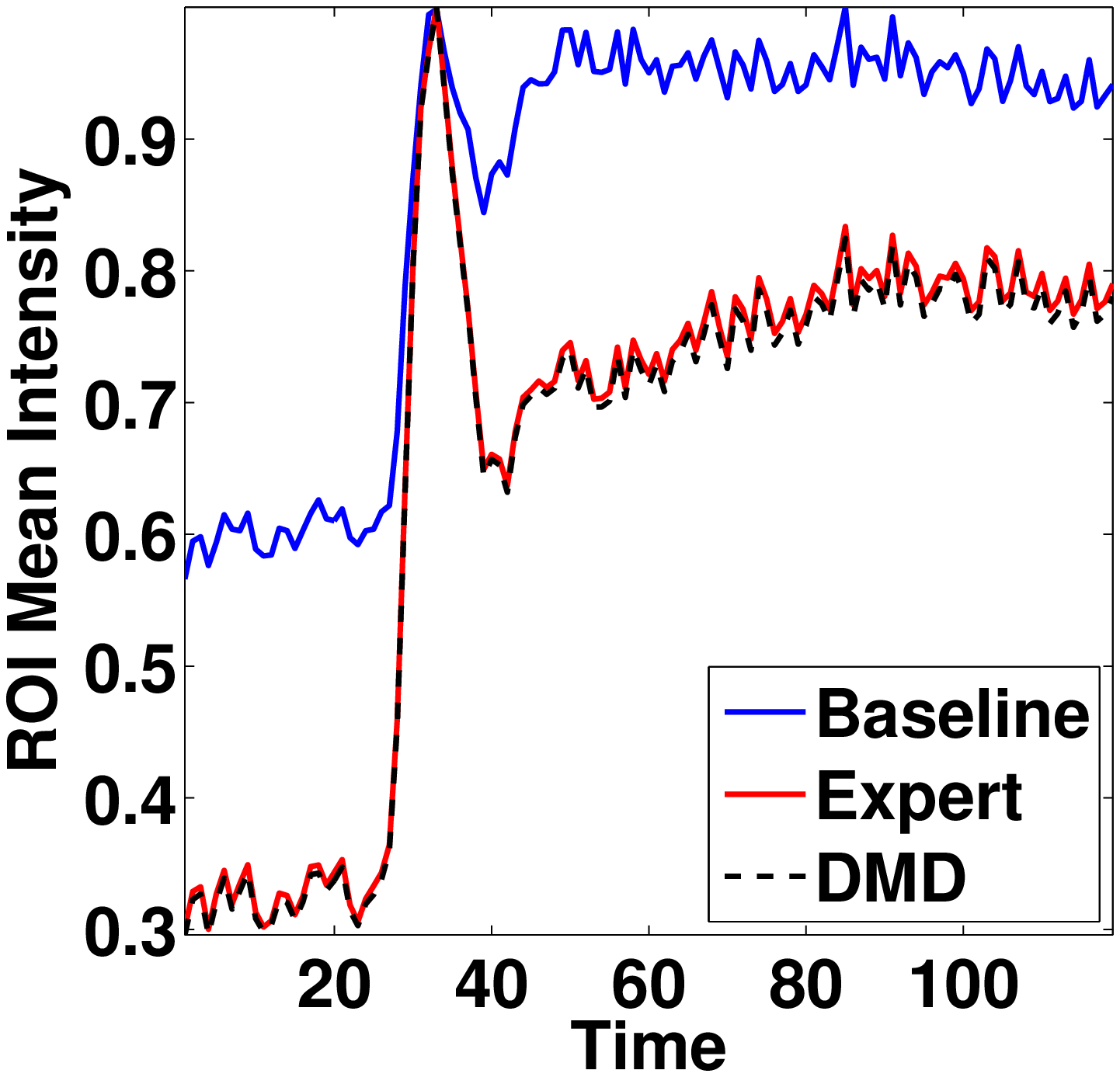} \\
(5) & (6) & (7) & (8) \\
& \includegraphics[scale=0.25]{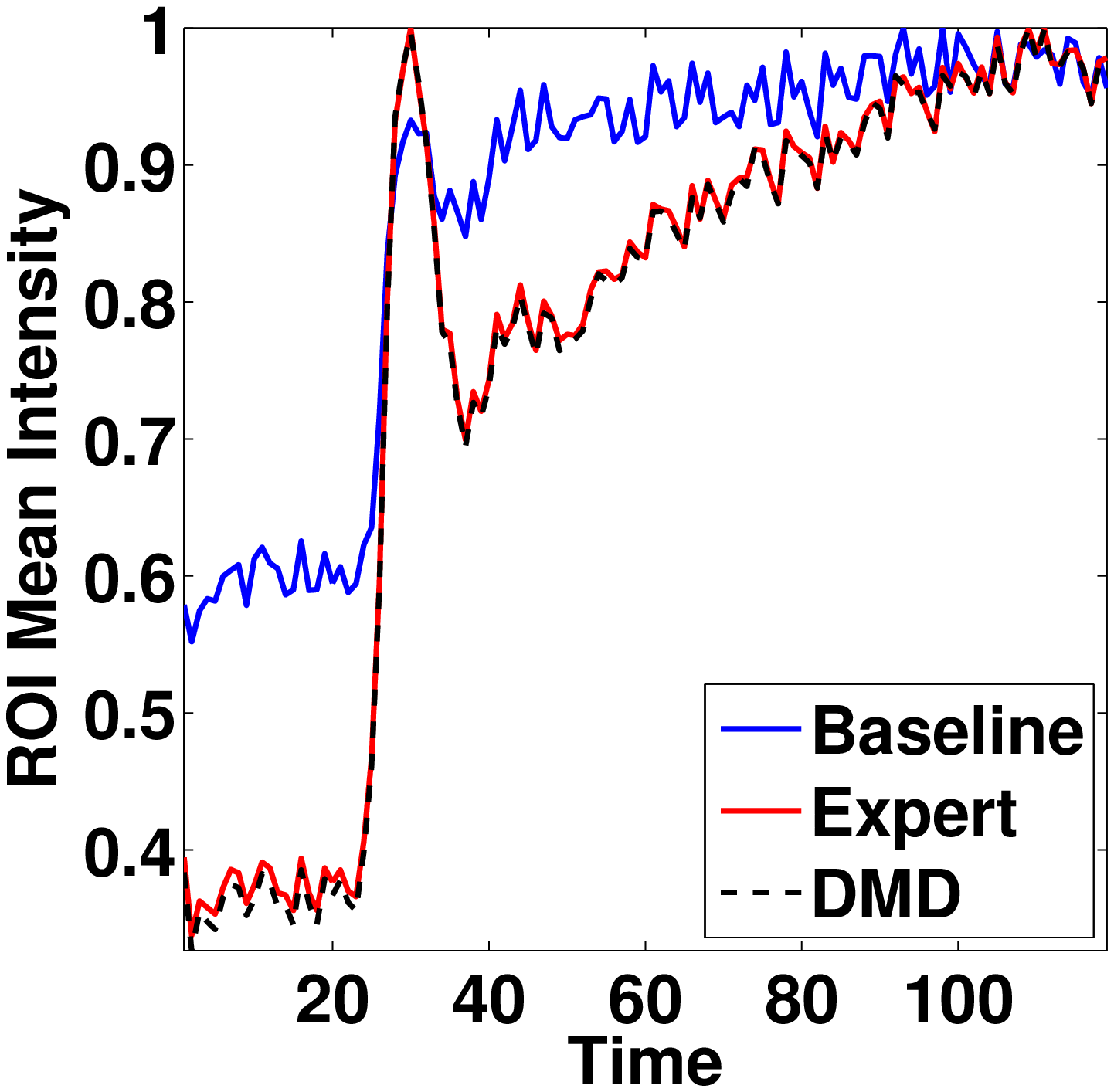}& \includegraphics[scale=0.25]{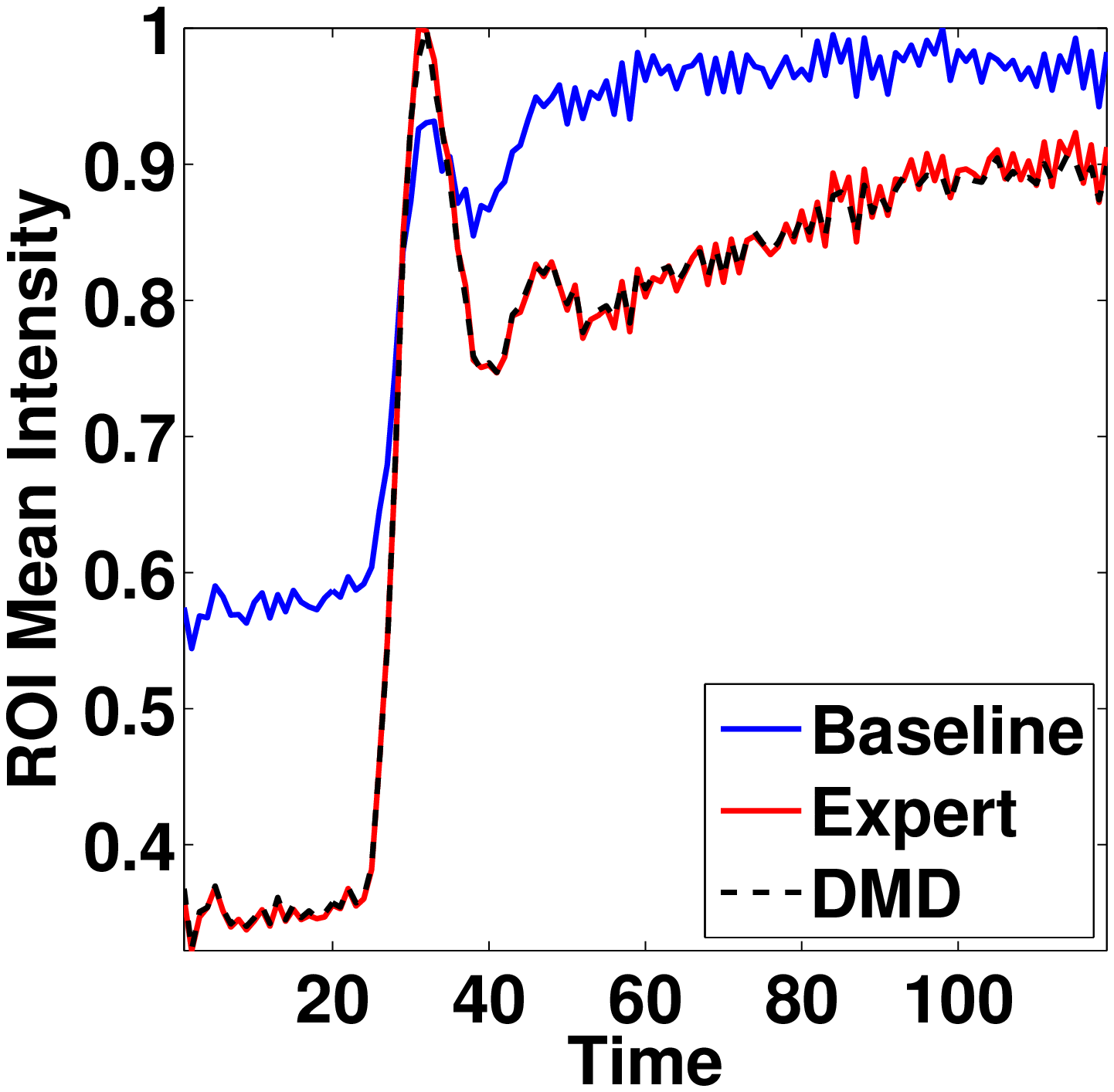} & \\
& (9) & (10) &  \\
\end{tabular}
\caption{Quantification of kidney region in DCE-MRI datasets 1 to 10.}
\label{result-1}
\end{figure*}

As a baseline method a minimum bounding box region around the kidney region (3) is selected as shown in Figure~\ref{fulspec-2}(a) across all the datasets. To evaluate the performance of the DMD framework we compare it against the performance of the manual delineation from a human expert. From Figure~\ref{result-1} we see that the kidney function produced from the automatic delineation clearly compares favourably with the human expert; where as the minimum bounding box region due to the influence of the PVE deviates from the kidney function quantified by a human expert. 

\begin{figure*}[h]
\begin{tabular}{c}
\includegraphics[scale=1]{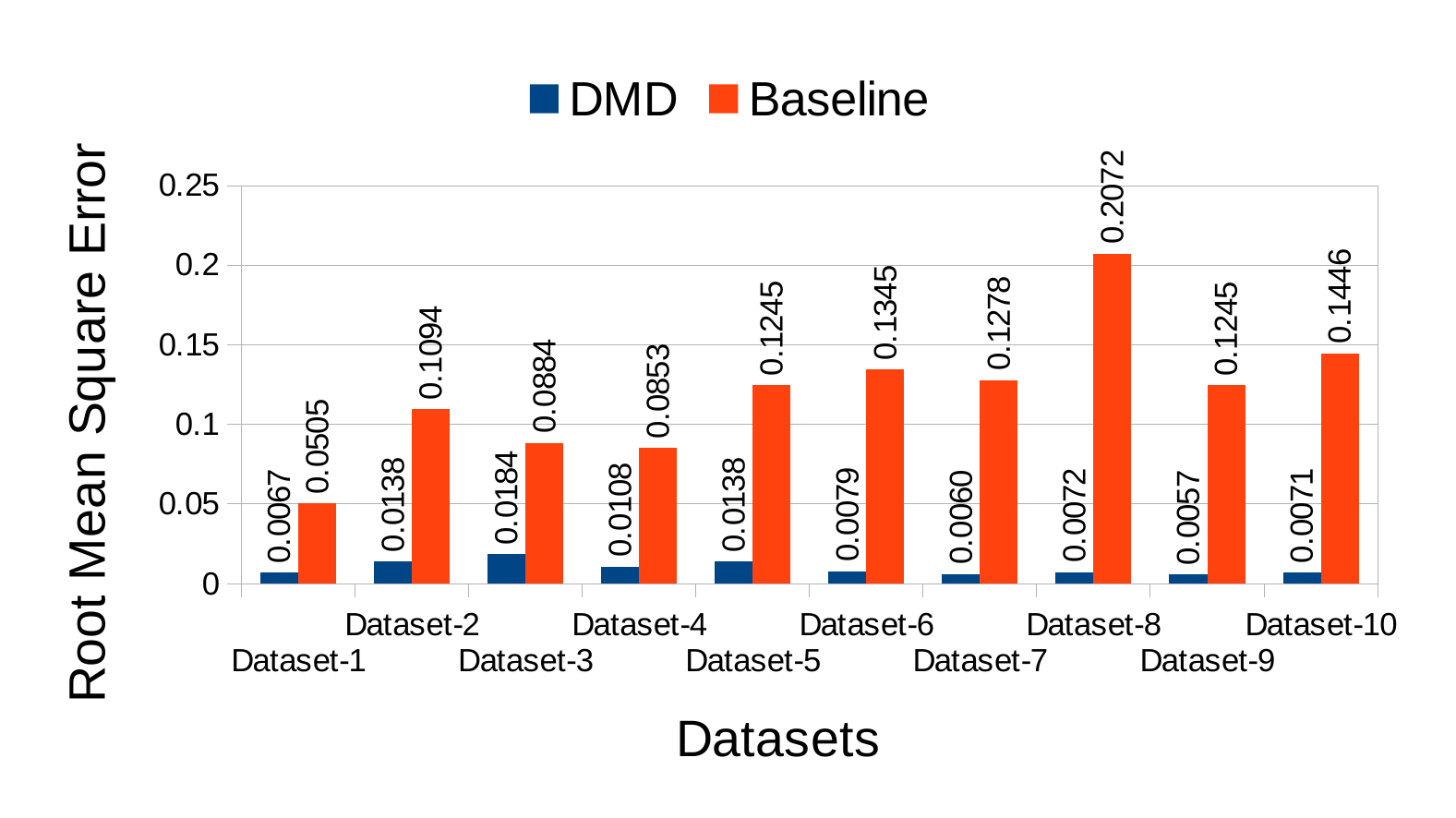}\\
\end{tabular}
\caption{RMSE for the time-intensity plots produced by DMD framework and baseline against human expert across 10 datasets.}
\label{result_2}
\end{figure*}

The root mean square error (RMSE) between the quantification produced by the proposed framework and  baseline method against the human expert is shown in Figure~\ref{result_2}. It is clear that the quantification results obtained from the proposed framework has lower RMSE when compared to the baseline method.  

\section{Conclusions}
\label{conc}

This study shows the significance of the proposed framework based on DMD, thresholding and blob analysis as a viable automatic delineation algorithm to effectively quantify the kidney function by reducing the partial volume effect in the DCE-MRI data. 

Our proposed DMD framework is applied on the DCE-MRI datasets collected from 10 healthy volunteers as well as the synthetic data mimicking the DCE-MRI data. We found that DMD can capture functional structures like kidney, liver and spleen in the top ranked dynamic modes. Particularly dynamic mode-2 capturing the kidney structure across all the datasets. This mode-2 is then thresholded and kidney template is obtained using connected component analysis. The kidney function produced from our proposed framework compares favourably with the human expert; while the minimum bounding box region due to the influence of the PVE deviates. Our results demonstrate that our proposed framework is very promising in delineating and quantifying the kidneys in DCE-MRI studies by removing the PVE.

\begin{acknowledgements}
The funding for this work has been provided by the Department of Computer Science and the Centre for Vision, Speech and Signal Processing (CVSSP) - University of Surrey. We also would like to express our gratitude towards Kidney Research UK for funding the DCE-MRI data acquisition as part of a reproducibility study.
\end{acknowledgements}

\bibliographystyle{spmpsci}      
\bibliography{myPaper,DMD_ref,myPaper_ckd,santosh_references,confirmation,movement_correction,SBI}
%
%

\end{document}